\pdfoutput=1
\documentclass[10pt,twocolumn,twoside]{IEEEtran}
\renewcommand{\baselinestretch}{1.06}
\usepackage{color}
\usepackage{theorem}
\usepackage{times,amsmath,epsfig}
\usepackage{amssymb}
\usepackage{cite}
\usepackage{algorithmic}
\usepackage{algorithm}
\usepackage[font=small,format=plain,up,up]{caption}
\usepackage{subcaption}
\usepackage{needspace}

\usepackage{tikz}
\usepackage{multicol}
\usetikzlibrary{shapes,arrows}
\input{mysymbol.sty}
\tikzstyle{phantom vertex} = [ ellipse, 
                               anchor = center, 
                               minimum height = 1*\unit, 
                               minimum width  = 1*\unit,
                               inner sep=0pt,
                               anchor=center]
\tikzstyle{red vertex}   = [black, fill = red!20,   phantom vertex, draw]
\tikzstyle{black vertex} = [black, fill = black!20, phantom vertex, draw]
\tikzstyle{blue vertex}  = [black, fill = blue!20,  phantom vertex, draw]
\tikzstyle{green vertex} = [black, fill = green!20,  phantom vertex, draw]
\tikzstyle{yellow vertex} = [black, fill = yellow!20,  phantom vertex, draw]
\tikzstyle{cyan vertex} = [black, fill = cyan!20,  phantom vertex, draw]
\tikzstyle{vertex}       = [draw, phantom vertex]

\tikzstyle{point} = [ellipse, inner sep=0pt, draw, fill=white, anchor = center,
                     minimum height = 0.05*\unit, minimum width  = 0.05*\unit]

\newcommand{\EE}{{\mathbb E}}

\newcommand{\Nv}{\mathrm{nv}}
\newcommand{\spanbasis}{\mathrm{span}}

\DeclareMathOperator{\Tr}{Trace}

\newtheorem{mycorollary}{\bf Corollary}
\newtheorem{mylemma}{\bf Lemma}
\newtheorem{myproposition}{\bf Proposition}
\newtheorem{remark}{\bf Remark}

\title{ Distributed Linear Network Operators using Graph Filters$^*$}

\author{\IEEEauthorblockN{Santiago Segarra, Antonio G. Marques, and Alejandro Ribeiro}
\thanks{$^*$This paper has been accepted for publication in the \textit{IEEE Transactions on Signal Processing} under the title ``Optimal Graph-Filter Design and Applications to Distributed Linear Network Operators''.}
\thanks{Work in this paper is supported {by the USA NSF CCF-1217963 and the Spanish MINECO grants No TEC2013-41604-R and TEC2016-75361-R.} S. Segarra is with the Inst. for Data, Systems and Society, Massachusetts Inst. of Technology. A. G. Marques is with the Dept. of Signal Theory and Comms., King Juan Carlos Univ. A. Ribeiro is with the Dept. of Electrical and Systems Eng., Univ. of Pennsylvania.  Emails: segarra@mit.edu, antonio.garcia.marques@urjc.es, and aribeiro@seas.upenn.edu. Part of the results in this paper were presented at the \textit{Allerton Conf.} \cite{ssamar_distfilters_allerton15} and ICASSP~\cite{ssamar_distfilters_icassp16}.}}

\begin{document}
\maketitle

%\markboth{IEEE TRANSACTIONS ON SIGNAL PROCESSING (ACCEPTED)}
%{SEGARRA \MakeLowercase{\textit{et al.}}: Optimal Graph-Filter Design and Applications to Distributed Linear Network Operators}

%%%%%%%%%%%%%%%%
\begin{abstract}%
%%%%%%%%%%%%%%%%
We study the optimal design of graph filters (GFs) to implement arbitrary linear transformations between graph signals. GFs can be represented by matrix polynomials of the graph-shift operator (GSO). {Since this operator captures the local structure of the graph, GFs naturally give rise to distributed operators. In most setups the GSO is given,} so that GF design consists fundamentally in choosing the (filter) coefficients of the matrix polynomial to resemble desired linear transformations. We determine spectral conditions under which a specific linear transformation can be implemented perfectly using GFs. For the cases where perfect implementation is infeasible, we address the optimization of the filter coefficients to approximate the desired transformation. Additionally, for settings where the GSO itself can be modified, we study its optimal design too. After this, we introduce the notion of a \emph{node-variant} GF, which allows the simultaneous implementation of multiple (regular) GFs in different nodes of the graph. This additional flexibility enables the design of more general operators without undermining the locality in implementation. Perfect and approximate designs are also studied for this new type of GFs. {To showcase the relevance of the results in the context of distributed linear network operators, the paper {closes} with the application of our framework to two particular distributed problems}: finite-time consensus and analog network coding.
%capitalizes on the fact that GFs can be implemented distributedly to 

%Finally, we present additional numerical experiments comparing the performance of node-invariant and node-variant filters when approximating arbitrary linear network operators.

\end{abstract}

%%%%%%%%%%%%%%%%
%\begin{keywords}
%Graph signal processing, Graph filter design, Distributed linear transformation, Network operator, Finite-time consensus, Analog network coding.
%\end{keywords}

{\bf \small \emph{Index terms} --- Graph signal processing, Graph filter design, Distributed linear transformation, Network operator, Finite-time consensus, Analog network coding.}

%%%%%%%%%%%%%%%%%%%%%%%%%%%%%%%%%%%%%%%%%%%
\section{Introduction}\label{S:Introduction}
%%%%%%%%%%%%%%%%%%%%%%%%%%%%%%%%%%%%%%%%%%%

%Networks and graphs can be understood as structures that encode pairwise relationships between elements of a set. 
Networks have often intrinsic value and are themselves the object of study. In other occasions, the network defines an underlying notion of proximity or dependence, but the object of interest is a signal defined on top of the graph, i.e., data associated with the nodes of the network. This is the matter addressed in the field of Graph Signal Processing (GSP), where notions such as frequency analysis, sampling, and stationarity are extended to signals supported on graphs \cite{EmergingFieldGSP,SandryMouraSPG_TSP13,SamplingKovacevic_without_Moura_15,marques2016stationaryTSP16}. {Graph-supported signals exist in a number of fields, with examples encompassing} gene-expression patterns defined on top of gene networks, the spread of epidemics over a social network, or the congestion level at the nodes of a telecommunication network. Transversal to the particular application, one must address the question of how to redesign traditional tools originally conceived to study and \emph{process} signals defined on regular domains and extend them to the more complex graph domain. Motivated by this, the present manuscript looks at the problem of optimal design of graph filters (GFs), and its applications in the context of distributed linear network operators.

Design of GFs to approximate a given \textit{impulse or frequency response} has been recently investigated in \cite{SandryMouraSPG_TSP13,isufi2016separable,isufi2017autoregressive}, considering both moving average and autoregressive configurations. This paper goes one step further and addresses the optimization of GFs to approximate a pre-specified \textit{generic} linear transformation. Mathematically, GFs are graph-signal operators that can be expressed as polynomials of the graph-shift~\cite{SandryMouraSPG_TSP14Freq}, which is a sparse matrix accounting for the local structure of the graph. Hence, GF design amounts to selecting the filter coefficients and, in certain applications, the entries of the graph-shift operator itself \cite{segarra2016topologyid}. An attractive feature of GFs is that they naturally give rise to (local) {\emph{distributed implementations}}. Therefore, our results are also relevant from the point of view of \emph{network} linear transformations. During the last decade, development of distributed linear schemes has attracted a lot of attention. The field has evolved quickly, shifting from the estimation of simple network parameters such as the mean using consensus algorithms~\cite{bertsekas1989parallelBook,XiaoBoyd2004,kar2013distributed,schizas2008consensus, AliConsensus08}, to advanced inference problems such as sparse linear regression~\cite{mateos2010dlasso}; see also~\cite{IntroSPMag_2013_Nets} for a review. Indeed, in the context of GSP but without leveraging the particular structure of a GF, several works have addressed the problem of \emph{distributed} implementation of different linear operators. Examples include the projection of graph signals onto low-dimensional spaces \cite{BarbarossaIcassp09}; the approximation of a specific class of linear transformations (the so-called graph Fourier multipliers) using Chebyshev polynomials \cite{ShumFrossard_ChebyshevDist2011}; or graph-signal inpainting \cite{wang2015local,Kovac_Dist_InpaintingICASPP15}. GFs have been used to implement distributedly \emph{specific} linear transformations such as fast average consensus \cite{sandryhaila2014finite}, projections onto the low-rank space of graph-bandlimited signals \cite{SuccNullingEigenv}, or interpolation of graph-bandlimited signals \cite{EUSIPCO_our_interp_2015, segarra2015reconstruction}, but not for general linear transformations. Design of distributed linear network operators under a GF framework is useful not only {because it identifies conditions and} rules for perfect and approximate implementation, but also because it allows us to leverage many of the insights and results existing for classical linear time-invariant filters.

%To describe our contribution more clearly, consider that each node has a certain value, and these values are collected across nodes to form a graph signal. With this definition, GFs are a specific class of operators whose input and output are graph signals. Mathematically, GFs are linear transformations that can be expressed as polynomials of the graph-shift operator~\cite{SandryMouraSPG_TSP14Freq}. The graph-shift operator can be viewed as a building block for more complex operators and, for the directed chain graph, it boils down to the classical time-shift operator. Although most existing works use either the Laplacian or the adjacency matrices as shift operators, any sparse matrix that encodes the structure of the underlying network can serve as a shift \cite{Oursampling_journal_2015}. For any given shift, the polynomial coefficients determine completely the GF and are referred to as \emph{filter coefficients}. The goal of this paper is to design the filter coefficients that, given a graph-shift operator, yield the best approximation to a pre-specified linear transformation.

%Hence, the problem of finding the filter coefficients (and possibly the shift) to approximate a given linear transformation is relevant not only from a theoretical point of view, but also from an application perspective. Since GFs act on graph signals through the successive application of local operators, 
To further motivate the practical interest of the investigated problem, one can exploit the fact that the output of a GF can be viewed as the outcome of a diffusion  or spreading process, with the filter coefficients corresponding to the rate of diffusion. Therefore, our results can be applied to network processes evolving according to a dynamics given by a (weighted) connectivity graph. The input of the GF is the initial network state, the output is the final network state, and the shift operator is given by the connectivity graph. The objective is then to design the filter coefficients, which amounts to tuning the parameters that control the dynamics, to approximate a given linear transformation. Another relevant example is distributed state estimation in power or communication networks, such as wireless sensor networks, {where lack of infrastructure or privacy issues prevents the deployment of a central controller and} the processing and exchange of information must be implemented locally. In this case, the entries of the shift correspond to the weights that each node gives to the information received from its neighbors, and the filter coefficients correspond to the weights that nodes give to the signals exchanged at different time instants.

The contributions and organization of the paper are as follows. In Section \ref{S:Modeling}, we first introduce the concepts of graph signals and (node-invariant) GFs, and then propose {two new types of GF,} whose coefficients are allowed to vary across nodes. Upon {casting} the problem of approximating linear \emph{network transformations} as a \emph{GF design}, Sections \ref{S:Impl_node_invar} and \ref{S:Impl_node_var} investigate perfect and approximate GF implementation. In Section \ref{S:Impl_node_invar}, the focus is on node-invariant GFs. First, conditions under which a specific linear transformation can be implemented perfectly are identified. These conditions depend on the spectrum of the linear transformation, the spectrum of the graph-shift operator, and the order of the filter. For the cases where the previous conditions are not met, the design of optimal filter coefficients to approximate the desired transformation under different error metrics is addressed. Moreover, for setups where the entries of the shift can be modified, the spectral conditions are also leveraged to design shifts that facilitate the implementation of the given transformation.  Section \ref{S:Impl_node_var} mimics the structure of Section \ref{S:Impl_node_invar}, but focusing on node-variant filters. To {illustrate} the relevance of our results for the {\emph{design of distributed network operators}}, Section~\ref{S:distributed_network_operators} applies the developed framework to the problems of finite-time consensus \cite{finiteconsensusKibangou11, sandryhaila2014finite} and analog network coding (ANC) \cite{Kattietal07ANC, Zhangetal06ANC}. Numerical experiments and concluding remarks in Sections~\ref{S:NumExper} and \ref{S:Concl} wrap-up the manuscript. {Relative to our conference papers \cite{ssamar_distfilters_allerton15} and  \cite{ssamar_distfilters_icassp16}, here we present a new definition for node-variant GF, provide proofs for our theoretical claims, investigate the problem of optimal shift design, discuss additional insights and generalizations, and report new numerical results.} 

\noindent \textbf{Notation:} In general, the entries of a matrix $\bbX$ and a vector $\bbx$ will be denoted as $X_{ij}$ and $x_i$; however, when contributing to avoid confusion, the alternative notation $[\bbX]_{ij}$ and $[\bbx]_i$ will be used. The notation $^T$, $^H$, and $^\dag$ stands for transpose, transpose conjugate, and Moore-Penrose pseudoinverse, respectively; $\lambda_{\max}(\bbX)$ is the largest eigenvalue of the symmetric matrix $\bbX$; $\bbe_i$ is the $i$-th $N\times 1$ canonical basis vector (all entries of $\bbe_i$ are zero except for the $i$-th one, which is one); $\mathbf{0}$ and $\mathbf{1}$ are the all-zero and all-one vectors, respectively; $\odot$ denotes the Khatri-Rao (column-wise Kronecker) product and $\circ$ the Hadamard (element-wise) product.

%\noindent \textbf{Notation:} Boldface capital letters denote matrices and boldface lowercase letters column vectors. Generically, the entries of a matrix $\bbX$ and a vector $\bbx$ will be denoted as $X_{ij}$ and $x_i$; however, when contributing to avoid confusion, the alternative notation $[\bbX]_{ij}$ and $[\bbx]_i$ will be used. The notation $^\dag$, $^T$ and $^H$ stands for conjugate, transpose and transpose conjugate, respectively; $\Tr(\bbX):=\sum_{\forall i} X_{ii}$ is the trace of the square matrix $\bbX$; $\lambda^{\max}(\bbX)$ is the largest eigenvalue of the symmetric matrix $\bbX$; $\diag(\bbx)$ is a diagonal matrix satisfying $[\diag(\bbx)]_{ii}=[\bbx]_i$; $\bbe_i$ is the $i$-th $N\times 1$ canonical vector (all entries of $\bbe_i$ are zero except the $i$-th one, which is one); $\mathbf{E}_K:=[\bbe_1,...,\bbe_K]$ is a tall matrix collecting the $K$ first canonical vectors; and $\mathbf{0}$ and $\mathbf{1}$ are, respectively, the all-zero and all-ones matrices (when not clear from the context, a subscript indicating the dimensions will be used). The modulus (remainder) obtained after dividing $x$ by $N$ will be denoted as $\mymod(x)$.

%%%%%%%%%%%%%%%%%%%%%%%%%%%%%%%%%%%%%%%%%%%%
\section{Graph signals and filters}\label{S:Modeling}
%%%%%%%%%%%%%%%%%%%%%%%%%%%%%%%%%%%%%%%%%%%%x

Let $\mathcal{G}$ denote a directed graph with a set of $N$ nodes or vertices $\mathcal{N}$ and a set of links $\mathcal{E}$, such that if node $i$ is connected to $j$, then $(i,j)\in\mathcal{E}$. The (incoming) neighborhood of $i$ is defined as the set of nodes $\mathcal{N}_i = \{j \,| \, (j,i)\in\mathcal{E}\}$ connected to $i$. For any given graph we define the adjacency matrix $\bbA$ as a sparse $N\!\times\! N$ matrix with non-zero elements $A_{ji}$ if and only if $(i,j)\in\ccalE$. The value of $A_{ji}$ captures the strength of the connection from $i$ to $j$. The focus of this paper is not on analyzing $\mathcal{G}$, but graph signals defined on the set of nodes $\mathcal{N}$. Formally, each of these signals can be represented as a vector $\bbx=[x_1,...,x_N]^T \in  \mathbb{R}^N$ where the $i$-th element represents the value of the signal at node $i$ or, alternatively, as a function $f : \mathcal{N} \to \mathbb{R}$, defined on the vertices of the graph.

The graph $\mathcal{G}$ is endowed with a \emph{graph-shift operator} $\bbS$ \cite{SandryMouraSPG_TSP13,SandryMouraSPG_TSP14Freq}. The operator $\bbS$ is an $N\!\times \!N$ matrix whose entry $S_{ji}$ can be non-zero only if $i\!=\!j$ or if $(i,j)\in\mathcal{E}$. The sparsity pattern of $\bbS$ captures the local structure of $\ccalG$, but we make no specific assumptions on the values of the non-zero entries of $\bbS$. Choices for $\bbS$ are $\bbA$ \cite{SandryMouraSPG_TSP13,SandryMouraSPG_TSP14Freq}, the graph Laplacian $\bbL$ \cite{EmergingFieldGSP}, and their respective generalizations \cite{godsil2001algebraic}. The intuition behind $\bbS$ is to represent a linear transformation that can be computed \textit{locally} at the nodes of the graph. More rigorously, if $\bby$ is defined as $\bby\!=\!\bbS\bbx$, then node $i$ can compute $y_i$ provided that it has access to the value of $x_j$ at $j\in \mathcal{N}_i$. We assume henceforth that $\bbS$ is diagonalizable, so that there exists an $N\!\times\! N$ matrix $\bbV$ and an $N\!\times\! N$ diagonal matrix $\bbLambda$ such that $\bbS=\bbV\bbLambda\bbV^{-1}$.
%In particular, $\bbS$ is diagonalizable when it is normal, i.e., it satisfies $\bbS\bbS^H=\bbS^H\bbS$ where $\bbS^H$ denotes the conjugate transpose of $\bbS$. In that case we have that $\bbV$ is unitary, which implies $\bbV^{-1}=\bbV^{H}$, and leads to the decomposition $\bbS=\bbV\bbLambda\bbV^H$. 
When $\bbS$ is normal{, meaning that} $\bbS\bbS^H=\bbS^H\bbS$, not only $\bbS$ is diagonalizable but $\bbV^{-1}=\bbV^{H}$ is unitary.
%, which implies $\bbV^{-1}=\bbV^{H}$, and leads to the decomposition $\bbS=\bbV\bbLambda\bbV^H$.
Given a graph signal $\bbx$, we refer to $\widehat{\bbx}:=\bbV^{-1}\bbx$ as the frequency representation of $\bbx$ \cite{SandryMouraSPG_TSP14Freq}. 
%Whenever $\widehat{\bbx}$ is sparse, we say that $\bbx$ is bandlimited.

A linear graph signal operator is a transformation $\bbB: \reals^N \to \reals^N$ between graph signals. Since the transformation is linear and the input and output spaces are the same, $\bbB$ can be represented by a square $N\times N$ matrix.

%---------------------------------------------------
\subsection{{GFs} and distributed implementation}\label{Ss:graph_filters}

Here we define (node-invariant) GFs, provide some intuition on their behavior, and discuss their distributed implementation. Mathematically, {with  $\bbc:=[c_0,\ldots,c_{L-1}]^T$ representing the vector of coefficients,} GFs are linear graph-signal operators $\bbH:\;\mathbb{R}^N \to \mathbb{R}^N$ of the form
\begin{equation}\label{E:def_graph_filters}
\bbH:=\sum_{l=0}^{L-1}c_l \bbS^l.
\end{equation}
{That is, {GFs are} \emph{polynomials} of the graph-shift operator \cite{SandryMouraSPG_TSP13}.} To emphasize its dependence on $\bbc$ and $\bbS$, the GF in \eqref{E:def_graph_filters} will be occasionally written as $\bbH(\bbc,\bbS)$. Leveraging the spectral decomposition of the shift, the GF $\bbH$ can also be written as $\bbH=\bbV\big(\sum_{l=0}^{L-1}c_l \boldsymbol{\Lambda}^l\big) \bbV^{-1}$. The diagonal matrix $\widehat{\bbH}:=\sum_{l=0}^{L-1}c_l \boldsymbol{\Lambda}^l$ can then be viewed as the frequency response of $\bbH$ and it can be alternatively written as $\widehat{\bbH}=\diag{(\widehat{\bbc})}$, where vector $\widehat{\bbc}$ is a vector that contains the $N$ frequency responses of the filter. Let $\lambda_k$ denote the $k$-th eigenvalue of $\bbS$ and define the $N \times L$ Vandermonde matrix
\begin{equation}\label{E:def_Psi_Vander}
\bbPsi:= \left( \begin{array}{cccc}
1 & \lambda_1 & \ldots & \lambda_1^{L-1} \\
\vdots & \vdots &  &\vdots\\
1 & \lambda_N &  \ldots  & \lambda_N^{L-1} \end{array} \right).
\end{equation}
For future reference, $D$ denotes the number of distinct eigenvalues in $\{\lambda_k\}_{k=1}^N$. Using \eqref{E:def_Psi_Vander}, it holds that $\widehat{\bbc}=\bbPsi\bbc$ and thus
\begin{equation}\label{E:Filter_from_time_to_freq}
\bbH=\!{ \sum_{l=0}^{L-1}c_l \bbS^l}=\!\bbV\diag\big(\bbPsi\bbc\big) \!\bbV^{-1}\!=\!\bbV\diag(\widehat{\bbc})\bbV^{-1}\!.
\end{equation}
This implies that if $\bby$ is defined as $\bby=\bbH\bbx$, its frequency representation $\widehat{\bby} = \bbV^{-1} \bby$ satisfies
\begin{equation}\label{E:Filter_input_output_freq}
\widehat{\bby}=\diag\big(\bbPsi\bbc\big)\bbV^{-1}\bbx=\diag\big(\widehat{\bbc}\big)\widehat{\bbx}=\widehat{\bbc}\circ\widehat{\bbx},
\end{equation}
which {shows} that the output at a given frequency depends only on the value of the input and the filter response at that given frequency. {In contrast with traditional discrete signal processing, where} the operator that transforms the signals and the filter coefficients into the frequency domain is the same -- the Discrete Fourier Transform (DFT) --, when a generic $\bbS$ is considered, matrices $\bbV^{-1}$ and $\bbPsi$ are different.

A particularity of GFs is that they can be implemented locally, e.g., with $L-1$ exchanges of information among neighbors. This is not surprising, since the shift $\bbS$ is a local operator. To formalize this, let us define the $l$-th shifted input signal as $\bbz^{(l)}:=\bbS^l\bbx$ and note that two properties of $\bbz^{(l)}$ are: i) it can be computed recursively (sequentially) as $\bbz^{(l)}=\bbS\bbz^{(l-1)}$, with $\bbz^{(0)}=\bbx$; and ii) node $i$ can obtain $[\bbz^{(l)}]_i$ locally based on the values of $[\bbz^{(l-1)}]_j$ at $j\in \mathcal{N}_i$. To emphasize this locality, we define $\bbz_i$ as an $L\times 1$ vector collecting the entries of $\{\bbz^{(l)}\}_{l=0}^{L-1}$ that are known by node $i$, so that $[\bbz_i]_l:=[\bbz^{(l)}]_i$.
With $\bby$ denoting the output of a GF for the input signal $\bbx$, it follows from \eqref{E:def_graph_filters} that
\begin{equation}\label{E:recursive_comp_filter}
\bby=\bbH \, \bbx=\sum_{l=0}^{L-1}c_l\bbS^l\bbx=\sum_{l=0}^{L-1}c_l\bbz^{(l)}.
\end{equation}
Hence, the $i$-th entry of vector $\bby$ can be computed as $y_i=\sum_{l=0}^{L-1}c_l[\bbz^{(l)}]_i=\bbc^T\bbz_i$, showing that if the nodes know the value of the filter coefficients, $y_i$ can be computed using solely information available at node $i$. This also suggests a two-step distributed implementation of GFs. First, $L-1$ sequential shifts are applied to the input signal $\bbx$ using only local information. At each shift, every node stores its own value of the shifted input. In the second step, nodes perform a linear combination of the values obtained in the first step. The weights, that are the same across nodes, are given by the coefficients $\bbc$. Alternatively, these two steps can be implemented simultaneously via a ``multiply and accumulate'' operation. A different expression to define a GF is
\begin{equation}\label{E:graph_filter_as_product}
\bbH= a_0 \prod_{l=1}^{L-1}(\bbS-a_l\bbI),
\end{equation}
which also gives rise to a polynomial on $\bbS$ of degree $L-1$. The previous expression can be leveraged to obtain an alternative distributed implementation of GFs. To show this, we define the intermediate graph signal $\bbw^{(l)}= \bbS\bbw^{(l-1)} - a_l \bbw^{(l-1)}$, with $\bbw^{(0)}=\bbx$. Notice that: i) the computation of $\bbw^{(l)}$ based on $\bbw^{(l-1)}$ can be performed by each node locally; and ii) the output of applying $\bbH$ to input $\bbx$ can be obtained after $L-1$ shifts just by setting $\bby=\bbH\bbx= a_0 \bbw^{(L-1)}$.

A convenient property of the representation in \eqref{E:graph_filter_as_product} is that it provides a straightforward way to design band-pass (frequency annihilating) filters \cite{SuccNullingEigenv}. In particular, upon setting $a_l = \lambda_k$ for the $k$-th eigenvalue of $\bbS$, filter $\bbH$ will eliminate the frequency basis $\bbv_k$, i.e., the eigenvector associated with $\lambda_k$. 
%It is also useful for shift operators with a large condition number. In such a case, the sequential implementation that first computes all $\{\bbz^{(l)}\}_{l=0}^{L-1}$ and then combines them linearly can give rise to numerical problems for high powers of $\bbS$. The implementation based on $\{\bbw^{(l)}\}_{l=0}^{L-1}$ is more robust to this type of problems by performing the linear combination and the application of the shift jointly.

\subsection{Node-variant GFs}\label{Ss:nodevariant_filters}

This paper proposes a generalization of GFs, called \emph{node-variant} GFs, as operators $\bbH_\Nv \!: \!\mathbb{R}^N \! \to \! \mathbb{R}^N$ of the form [cf.~\eqref{E:def_graph_filters}]
\begin{equation}\label{E:node_variant_filter}
\bbH_{\Nv}:= \sum_{l=0}^{L-1}\diag (\bbc^{(l)}) \bbS^l.
\end{equation}
Whenever the $N \times 1$ vectors $\bbc^{(l)}$ are constant, i.e. $\bbc^{(l)} = c_l \mathbf{1}$ for all $l$, the node-variant GF reduces to the regular (node-invariant) GF. However, for general $\bbc^{(l)}$, when $\bbH_\Nv$ is applied to a signal $\bbx$, each node applies different weights to the shifted signals $\bbz^{(l)} = \bbS^l \bbx$. This additional flexibility enables the design of more general operators without undermining the local implementation. For notational convenience, the filter coefficients associated with node $i$ are collected in the $L\times 1$ vector $\bbc_i$, such that $[\bbc_i]_l=[\bbc^{(l)}]_i$, and we define the $L \times N$ matrix $\bbC := [\bbc_1, \ldots, \bbc_N]$. The GFs in \eqref{E:node_variant_filter} can be viewed as a generalization of linear time-varying filters whose impulse response changes with time. Whenever convenient to emphasize their dependence on the filter coefficients or the shift, the GF $\bbH_{\Nv}$ in \eqref{E:node_variant_filter}  will be written as $\bbH_{\Nv}\big(\{\bbc_i\}_{i=1}^{N},\bbS\big)$.

Since $\bbS^l$ and $\diag(\bbc^{(l)})$ are not simultaneously diagonalizable (i.e., their eigenvectors are not the same), the neat frequency interpretation succeeding \eqref{E:def_graph_filters} does not hold true for the filters in \eqref{E:node_variant_filter}. However, the spectral decomposition of $\bbS$ can still be used to understand how the output of the filter at a given node $i$ depends on the frequency components of the input.
To be specific, let us write the filter in \eqref{E:node_variant_filter} as
\begin{equation}\label{E:input_output}
\bbH_{\Nv}=\sum_{l=0}^{L-1}\diag (\bbc^{(l)}) \bbV\boldsymbol{\Lambda}^l \bbV^{-1}.
\end{equation}
Next, to analyze the effect of $\bbH_{\Nv}$ on the value of the output signal at node $i$, consider the $i$-th row of $\bbH_{\Nv}$, given by
\begin{equation}\label{E:row_H}
\bbh_{i}^T:=\bbe_i^T\bbH_{\Nv}=\sum_{l=0}^{L-1} [\bbc_i]_l \bbe_i^T\bbV\boldsymbol{\Lambda}^l \bbV^{-1},
\end{equation}
with $\bbe_i$ being the $i$-th $N\!\times\! 1$ canonical vector. Defining the vectors $\bbu_i:=\bbV^T\bbe_i$ and $\widehat{\bbc}_i:=\bbPsi\bbc_i$, we can rewrite \eqref{E:row_H} as
\begin{align}\label{E:row_H_v2_a}
\bbh_{i}^T& =\sum_{l=0}^{L-1} [\bbc_i]_l \bbu_i^T \boldsymbol{\Lambda}^l \bbV^{-1}=\bbu_i^T \Big(\sum_{l=0}^{L-1} [\bbc_i]_l  \boldsymbol{\Lambda}^l\Big) \bbV^{-1} \nonumber\\
&=\bbu_i^T\diag(\bbPsi\bbc_i)\bbV^{-1}=\bbu_i^T\diag(\widehat{\bbc}_i)\bbV^{-1}.
\end{align}
The expression in \eqref{E:row_H_v2_a} reveals that the output of the filter at node $i$, which is given by $\bbh_i^T\bbx$, can be viewed as an inner product of $\bbV^{-1}\bbx$ (the frequency representation of the input) and $\bbu_i$ (how strongly node $i$ senses each of the frequencies), modulated by $\widehat{\bbc}_i$ (the frequency response associated with the coefficients used by node $i$).
%On the other hand, the expression in \eqref{E:row_H_v2_b} can be used to design filter coefficients $\{\bbc_i\}_{i=1}^N$ that implement a given linear transformation $\bbB$.
The expressions in \eqref{E:row_H_v2_a} will be leveraged in Section \ref{S:Impl_node_var} to identify the type of linear transformations that node-variant filters are able to implement.

{{As a final remark, note that an alternative definition for node-variant GFs -- different from the one in \eqref{E:node_variant_filter} -- is to consider} 
\begin{equation}\label{E:node_variant_filter_typeII}
\bbH_{\Nv}':= \sum_{l=0}^{L-1} \bbS^l\diag (\bbc^{(l)}).
\end{equation}
As was the case for \eqref{E:node_variant_filter}, if $\bbc^{(l)}=c_l\bbone$, then \eqref{E:node_variant_filter_typeII} is equivalent to the node-invariant GF in \eqref{E:def_graph_filters}. For each of the terms in \eqref{E:node_variant_filter_typeII} the filter first modulates the signal $\bbx$ with $\bbc^{(l)}$ and then applies the shift $\bbS^l$. This is in contrast with \eqref{E:node_variant_filter}, which first shifts $\bbx$ and then modulates the shifted input with the corresponding $\bbc^{(l)}$. {As was the case for \eqref{E:node_variant_filter}, it is possible to implement \eqref{E:node_variant_filter_typeII} in a distributed and sequential fashion.} To be more specific, to obtain the output signal $\bby=\bbH_{\Nv}'\bbx$ one starts with the intermediate graph signal $\bbt^{(0)}:=\bbzero$, computes $\bbt^{(l)}\in \reals^N$ for $l>0$ as $\bbt^{(l)}:=\bbS\bbt^{(l-1)}+\diag(\bbc^{(L-l)})\bbx$, and sets $\bby=\bbt^{(L)}$. Although the rest of the paper focuses on node-variant GFs of the form in \eqref{E:node_variant_filter}, the frequency interpretations in \eqref{E:row_H} and \eqref{E:row_H_v2_a} as well as the optimal designs derived in the ensuing sections {can be generalized for the GFs in \eqref{E:node_variant_filter_typeII}.}}

\begin{remark}
\normalfont {Node-variant GFs come with some caveats, one of them being that their spectral interpretation, {which is a key} component in most graph signal processing works, is more involved than that of node-invariant GFs. On the other hand, they exhibit a number of advantages related to their flexibility. As already pointed out, from a practical point of view they offer a way to implement a very large class of linear operators in a distributed manner (see the discussions after \eqref{E:node_variant_filter} and \eqref{E:node_variant_filter_typeII}, and the results in Sections \ref{S:Impl_node_var} and \ref{S:distributed_network_operators}). {Moreover, just like classical time-varying filters, }node-varying GFs are expected to play an important role in problems related to adaptive graph filtering or modeling of non-stationary graph processes, as well as in establishing links with control theory and modeling of local diffusion dynamics.}
\end{remark}

%%%%%%%%%%%%%%%%%%%%%%%%%%%%%%%%%%%%%%%%%%%%
\section{Optimal design of node-invariant GFs}\label{S:Impl_node_invar}
%%%%%%%%%%%%%%%%%%%%%%%%%%%%%%%%%%%%%%%%%%%%

{The first goal} is to use the node-invariant GFs in \eqref{E:def_graph_filters} to implement a pre-specified \emph{linear} graph signal transformation $\bbB$. To be more concrete, we want to find the filter coefficients $\bbc$ such that the following equality
\begin{equation}\label{E:linear_transf_equal_filter}
\bbB=\bbH\big(\bbc,\bbS\big)=\sum_{l=0}^{L-1}c_l\bbS^l
\end{equation}
holds. This is addressed in Section~\ref{Ss:perfect_impl_node_invar}, while approximate solutions are analyzed in Section~\ref{Ss:approx_impl_node_invar}.

%
%Conditions under which this equality can be achieved are discussed in Section~\ref{Ss:perfect_impl_node_invar}. In Section~\ref{Ss:approx_impl_node_invar} we present results on imperfect reconstruction where the goal is to design the filter coefficients such that either the linear transformation obtained is close to $\bbB$ or the filtered signal is close to $\bbB \bbx$ for some statistical model of the input $\bbx$.

%---------------------------------------------------
\subsection{Conditions for perfect implementation}\label{Ss:perfect_impl_node_invar}
With $\bbbeta:=[\beta_1,...,\beta_N]^T$ {standing for the} vector containing the eigenvalues of matrix $\bbB$ (including multiplicities), the conditions for perfect implementation are given next.

%%%%%%%%% P R O P O S I T I O N %%%%%%%
\begin{myproposition}\label{P:Perfect_Impl_NodeInvarFilter}
The linear transformation $\bbB$ can be implemented using a node-invariant GF $\bbH=\bbH(\bbc,\bbS)$ of the form given in \eqref{E:def_graph_filters} if the three following conditions hold true:\\
\noindent a) Matrices $\bbB$ and $\bbS$ are simultaneously diagonalizable; i.e., all the eigenvectors of $\bbB$ and $\bbS$ coincide.\\
\noindent b) For all $(k_1,\!k_2)$ such that $\lambda_{k_1} = \lambda_{k_2}$, it holds that $\beta_{k_1} = \beta_{k_2}$. \\
\noindent c) The degree of $\bbH$ is such that $L\geq D$.
\end{myproposition}
%%%%%%
\begin{myproof}
Using condition \emph{a)}, we may reformulate the equivalence between $\bbB$ and $\bbH$ in the frequency domain and require their eigenvalues to be the same, i.e. $\bbbeta = \bbPsi \bbc$. Denote by $\ccalK$ the set of indices of the $D$ unique eigenvalues of $\bbS$ and define the $N \times D$ matrix $\bbE_{\ccalK}=[\bbe_{k_1}, \ldots, \bbe_{k_D}]$ for all $k_i \in \ccalK$ and the $N \times N \!-\! D$ matrix $\bbE_{\bar{\ccalK}}\!=\![\bbe_{k_1}, \ldots, \bbe_{k_{N-D}}]$ for all $k_i\! \not\in \!\ccalK$. Then, we split the system of equations $\bbbeta = \bbPsi \bbc$ into two groups
\begin{align}
\bbE_{\ccalK}^T \bbbeta = \bbE_{\ccalK}^T \bbPsi \bbc, \label{E:equivalence_eigenvalues_1} \\
\bbE_{\bar{\ccalK}}^T \bbbeta = \bbE_{\bar{\ccalK}}^T \bbPsi \bbc. \label{E:equivalence_eigenvalues_2}
\end{align}
Condition \emph{b)} guarantees that if $\bbc$ solves \eqref{E:equivalence_eigenvalues_1} then it also solves \eqref{E:equivalence_eigenvalues_2}. Finally, combining the fact that $\bbE_{\ccalK}^T \bbPsi$ has full row rank (Vandermonde matrix with non-repeated basis) and condition \emph{c)}, there exists at least one $\bbc$ that solves \eqref{E:equivalence_eigenvalues_1}, and the proof concludes.
\end{myproof}
%%%END OF THE PROOF OF THE PROPOSITION%%%

Based on the proposition, the expression for the optimal coefficients follows readily, as stated in the following corollary.
%which to improve readability is given in the form of a corollary.  
\begin{mycorollary}
If the conditions in Proposition~\ref{P:Perfect_Impl_NodeInvarFilter} hold true, a vector of filter coefficients $\bbc^*$ satisfying \eqref{E:linear_transf_equal_filter} is given by
\begin{equation}\label{E:syst_equations_to_solve_perfect_case}
\bbc^* = \bbPsi^\dagger\bbbeta.
%\bbbeta = \widehat{\bbc}=\bbPsi\bbc.
\end{equation}
\end{mycorollary}
\begin{myproof}
If the conditions in Proposition~\ref{P:Perfect_Impl_NodeInvarFilter} hold, then there exists a vector $\bbc$ such that the cost $\|\bbbeta-\bbPsi\bbc\|_2^2$ is zero. Since the pseudo-inverse solution $\bbc^*$ in \eqref{E:syst_equations_to_solve_perfect_case} is one of the solutions to the minimization of $\|\bbbeta-\bbPsi\bbc\|_2^2$, we have that $\|\bbbeta-\bbPsi\bbc^*\|_2^2=0$ and the corollary follows. When $L=D$, the solution is unique and, using the set $\ccalK$ defined before \eqref{E:equivalence_eigenvalues_1}, the expression in \eqref{E:syst_equations_to_solve_perfect_case} can be alternatively written as $\bbc^*=(\bbE_{\ccalK}^T \bbPsi)^{-1}(\bbE_{\ccalK}^T \bbbeta)$.
\end{myproof}

The corollary {leverages the spectral interpretation of GFs, advocating for a design where the filter coefficients are found in the frequency domain}. To be precise, if the conditions in Proposition \ref{P:Perfect_Impl_NodeInvarFilter} hold then: i) $\bbbeta$ 
%(which can be found either by computing the eigendecomposition of $\bbB$, or, by collecting the diagonal entries of $\bbV^{-1}\bbB\bbV$) 
represents the desired frequency response of the filter $\bbH=\sum_{l=0}^{L-1}c_l\bbS^l$; and ii) the pseudo-inverse $\bbPsi^\dagger$ in \eqref{E:syst_equations_to_solve_perfect_case} is the generalization of the \textit{inverse} graph Fourier transform when matrix $\bbPsi$ is not square. 
%Note also that while conditions \emph{a)} and \emph{b)} are necessary, the additional requirement in \emph{c)} is sufficient but not necessary for every $\bbB$, i.e., there exist linear transformations that can be implemented with filters of degree smaller than $D-1$. For example, implementing the transformation $\bbB = \bbI - \bbS$, which satisfies \emph{a)} and \emph{b)}, requires a GF of degree 1, regardless of the value of $D$.

To gain some intuition on the meaning of the proposition, let $\bbA_{dc}$ be the adjacency matrix of the directed cycle, which is the support of classical time-varying signals \cite{EmergingFieldGSP}. {When} Proposition \ref{P:Perfect_Impl_NodeInvarFilter} is particularized to $\bbS=\bbA_{dc}$, condition a) implies that {the transformations} $\bbB$ that can be implemented perfectly are those diagonalized by the DFT matrix, i.e., circulant transformations. In other words, if the output of the filter is a linear combination of successive applications of the shift $\bbA_{dc}$ to the input, only transformations that are time invariant (so that each row of $\bbB$ is the shifted version of the previous row) can be implemented perfectly using $\bbH(\bbc,\bbA_{dc})$. 

{Proposition \ref{P:Perfect_Impl_NodeInvarFilter} will} also be exploited in Section \ref{Ss:Designing_the_shift} to design shifts able to implement desired linear transformations.

%---------------------------------------------------
\subsection{Approximate implementation}\label{Ss:approx_impl_node_invar}

In general, if the conditions in Proposition~\ref{P:Perfect_Impl_NodeInvarFilter} are not satisfied, perfect implementation of $\bbB$ is not feasible. In such cases, the filter coefficients {$\bbc$} can be designed to minimize a pre-specified error metric. We suppose first that prior knowledge of the input signal $\bbx$ is available. This knowledge can be incorporated into the design {of $\bbc$} and the goal is then to minimize an error metric of the difference \emph{vector} $\bbd:=\bbH\bbx-\bbB\bbx$. A case of particular interest is when $\bbx$ is drawn from a zero-mean distribution with known covariance $\bbR_{\bbx}:=\mathbb{E}[\bbx\bbx^T]$. In this case, the error covariance is given by $\bbR_{\bbd}:=\EE[\bbd \bbd^T] = (\bbH-\bbB)\bbR_{\bbx}(\bbH-\bbB)^T$. Our objective is to pick $\bbc$ to minimize some metric of the error covariance matrix $\bbR_{\bbd}$. Two commonly used approaches are the minimization of $\Tr(\bbR_{\bbd})$ and $\lambda_{\max}(\bbR_{\bbd})$, where the former is equivalent to minimizing the mean squared error (MSE) of $\bbd$ and the latter minimizes the worst-case error (WCE) by minimizing the maximum variance of $\bbd$ \cite{pukelsheim1993optimal}.

Define the $N^2 \!\times\! L$ matrix $\bbTheta_{\bbR_{\bbx}} := [ \mathrm{vec}(\bbI \bbR_{\bbx}^{1/2}),$ $\,  \mathrm{vec}(\bbS \bbR_{\bbx}^{1/2}),$ $  \ldots, \mathrm{vec}(\bbS^{L-1} \bbR_{\bbx}^{1/2})]$ and the $N^2 \times 1$ vector $\bbb_{\bbR_{\bbx}}  := \mathrm{vec}(\bbB \bbR_{\bbx}^{1/2})$. The optimal filter coefficients are provided in the following proposition.
%
%PROPOSITION
\begin{myproposition}\label{P:Imperfect_Impl_NodeInvarFilter_2}
The optimal MSE filter coefficients $\bbc_\mathrm{Tr}^* := \argmin_{\bbc} \Tr(\bbR_{\bbd})$ are given by
\begin{equation}\label{E:optimal_filter_coefficients_case_2}
\bbc_\mathrm{Tr}^* = \bbTheta_{\bbR_{\bbx}}^\dag \bbb_{\bbR_{\bbx}} \stackrel{ \star}{=} (\bbTheta_{\bbR_{\bbx}}^T \bbTheta_{\bbR_{\bbx}})^{-1} \bbTheta_{\bbR_{\bbx}}^T \bbb_{\bbR_{\bbx}},
\end{equation}
where $\stackrel{\star}{=}$ holds if $\bbTheta_{\bbR_{\bbx}}$ has full column rank; while the optimal WCE coefficients $\bbc_{\lambda}^* := \argmin_{\bbc} \lambda_{\max}(\bbR_{\bbd})$ are obtained as
\begin{align}\label{E:optimal_filter_coefficients_case_2_2}
&\{ \bbc_\lambda^*, s^*\} \,\, = \,\, \argmin_{\{\bbc, s\}} \qquad s \\
&\mathrm{s.\,\,to} \,\,\, \begin{bmatrix}
s \bbI &  \bbV \diag(\bbPsi \bbc) \bbV^{-1} - \bbB \\
(\bbV\diag(\bbPsi \bbc) \bbV^{-1} - \bbB)^T & s \bbR_\bbx^{-1}
\end{bmatrix}
\!\! \succeq 0. \nonumber
\end{align}

\end{myproposition}
%%%%%%%%
\begin{myproof}
By expressing $\bbR_\bbx = \bbR_\bbx^{1/2} \bbR_\bbx^{1/2}$, we may rewrite $\Tr(\bbR_{\bbd})$ as $\Tr( (\bbH \bbR^{1/2}_\bbx - \bbB \bbR^{1/2}_\bbx) (\bbH \bbR^{1/2}_\bbx - \bbB \bbR^{1/2}_\bbx)^T )$. Thus, given that for any real matrix $\bbA$ we have that $\| \bbA \|_{\mathrm{F}}^2 = \Tr(\bbA \bbA^T)$, it follows that
\begin{equation}\label{E:proof_Imperfect_Impl_NodeInvarFilter_2_010}
\Tr(\bbR_\bbd) = \| \bbH \bbR^{1/2}_\bbx - \bbB \bbR^{1/2}_\bbx \|_\mathrm{F}^2.
\end{equation}
This implies that $\bbc_\mathrm{Tr}^*$ minimizes the Frobenius norm in \eqref{E:proof_Imperfect_Impl_NodeInvarFilter_2_010}.
From the definition of Frobenius norm, we have that $\| \bbH\bbR^{1/2}_\bbx - \bbB\bbR^{1/2}_\bbx \|_\mathrm{F} = \| \mathrm{vec}(\bbH\bbR^{1/2}_\bbx) - \mathrm{vec}(\bbB\bbR^{1/2}_\bbx)\|_2$. The result in \eqref{E:optimal_filter_coefficients_case_2} follows from noting that $\mathrm{vec}(\bbH\bbR^{1/2}_\bbx) = \bbTheta_{\bbR_{\bbx}} \bbc$ and using the pseudoinverse to solve the system of linear equations \cite{Golub_MatComp_Book}.

To show the result for $\bbc_\lambda^*$, note that the constraint in \eqref{E:optimal_filter_coefficients_case_2_2} can be rewritten in terms of its Schur complement as $s^2 \bbI \succeq (\bbH - \bbB) \bbR_\bbx (\bbH - \bbB)^T$. In terms of eigenvalues, this is equivalent to $s^2 \geq \lambda_{\max}(\bbR_{\bbd})$, from where it follows that $s^2$ is an upper bound on $\lambda_{\max}(\bbR_{\bbd})$. Since the constraint in \eqref{E:optimal_filter_coefficients_case_2_2} is tight, minimizing the upper bound is equivalent to minimizing $\lambda_{\max}(\bbR_{\bbd})$, so that the obtained coefficients $\bbc_\lambda^*$ are optimal, as stated in the proposition; see \cite{overton1988minimizing,boyd1993method}.
\end{myproof}
%%%END OF THE PROOF OF THE PROPOSITION%%%

Proposition~\ref{P:Imperfect_Impl_NodeInvarFilter_2} provides a closed-form expression for the MSE coefficients $\bbc_\mathrm{Tr}^*$ and obtains the WCE coefficients $\bbc_\lambda^*$ as the solution of a semi-definite convex program \cite{XiaoBoyd2004}. 
Whenever the spectrum of $\bbB$ is incompatible with that of $\bbS$ -- violations on conditions \emph{a)} or \emph{b)} in Proposition~\ref{P:Perfect_Impl_NodeInvarFilter} -- or when the filter degree is lower than that needed for perfect reconstruction, expressions \eqref{E:optimal_filter_coefficients_case_2} and \eqref{E:optimal_filter_coefficients_case_2_2} {become relevant}. Alternative error metrics and additional assumptions could be incorporated into the designs including, but not limited to, supplementary statistical knowledge of $\bbx$ and structural properties such as sparsity or bandlimitedness \cite{SandryMouraSPG_TSP14Freq}. For example, if $\bbS$ is normal and $\bbx$ is graph stationary \cite{marques2016stationaryTSP16}, then the designs in \eqref{E:optimal_filter_coefficients_case_2} and \eqref{E:optimal_filter_coefficients_case_2_2} can be addressed in the frequency domain, with the benefit of all involved matrices being diagonal. Finally, if no prior information on $\bbx$ is available, we may reformulate the optimal criterion to minimize the discrepancy between the \emph{matrices} $\bbH$ and $\bbB$. The optimal solutions are still given by \eqref{E:optimal_filter_coefficients_case_2} and \eqref{E:optimal_filter_coefficients_case_2_2} by setting $\bbR_{\bbx}=\bbI$. More precisely, the resulting coefficients $\bbc_{\rm Tr}^*$ in \eqref{E:optimal_filter_coefficients_case_2} minimize the error metric $\| \bbH - \bbB \|_{\mathrm{F}}$, while $\bbc_{\lambda}^*$ in \eqref{E:optimal_filter_coefficients_case_2_2} is the minimizer of $\| \bbH - \bbB \|_{2}$.

%\eqref{E:optimal_filter_coefficients_case_2} and \eqref{E:optimal_filter_coefficients_case_2_2} can be {solved with} $\bbR_{\bbx}=\bbI$. The resultant coefficients $\bbc_{\rm Tr}^*$ will then minimize the error metric $\| \bbH - \bbB \|_{\mathrm{F}}$, while $\bbc_{\lambda}^*$ will be the minimizer of $\| \bbH - \bbB \|_{2}$.

\subsection{Designing the shift}\label{Ss:Designing_the_shift}

In this section we go beyond the design of the \textit{filter coefficients} and discuss how to use the results in Proposition \ref{P:Perfect_Impl_NodeInvarFilter} to design a \textit{shift} $\bbS$ that enables the implementation of a given linear transformation $\bbB$.  
	
Since Proposition \ref{P:Perfect_Impl_NodeInvarFilter} established that $\bbB$ and $\bbS$ need to share the entire set of eigenvectors, we first look at the case where $\bbB$ is a rank-one matrix that can be written as $\bbB_{\mathrm{rk1}} := \mathbf{a}\mathbf{b}^T$. Finding a shift able to implement $\bbB=\bbB_{\mathrm{rk1}}$ is a favorable setup because $\bbB_{\mathrm{rk1}}$ specifies only one of the eigenvectors of $\bbS$, while the others can be chosen to endow $\bbS$ with desired properties. Rank-one operators are common in, for example, distributed estimation setups such as a wireless sensor network, where the goal is for each of the sensors to obtain the same estimate. To be specific, suppose that the estimate $\hat{\theta}$ can be written as a linear combination of the observations available at the nodes, i.e., $\hat{\theta}=\sum_{i=1}^N b_ix_i$ where $\bbx=[x_1,...,x_N]^T$ are the observations and $\bbb=[b_1,...,b_N]^T$ the weighting coefficients. Then, the sought network transformation is $\bbB_{\mathrm{rk1}}=\bbone \bbb^T$, so that when $\bbB_{\mathrm{rk1}}$ is applied to $\bbx$ gives rise to $\hat{\theta}\bbone$. 

The fact that $N-1$ eigenvalues of $\bbB_{\mathrm{rk1}}=\bba \bbb^T$ are equal to zero can be leveraged to show the following proposition.

%%%%%%%%%% COROLLARY
\begin{myproposition}\label{P:shift_rankone}
	Consider the graph $\ccalG$ and the rank-one transformation $\bbB_{\mathrm{rk1}}= \bba \bbb^T$. If $\ccalG$ is connected and $a_i\neq0$ and $b_i\neq0$ for all $i$, then there exists a shift $\bbS$ associated with $\ccalG$ such that the transformation $\bbB_{\mathrm{rk1}}$ can be written as a node-invariant GF $\bbH(\bbc,\bbS)$ of the form given in \eqref{E:def_graph_filters}.
\end{myproposition}
%%%%%%%
\begin{myproof}
The proof is constructive. Suppose without loss of generality that $\bba$ and $\bbb$ have unit norm and let $\ccalT$ denote a spanning tree of $\ccalG$ with edge set $\ccalE_{\ccalT}$. With $\mu\in \reals$ being an arbitrary constant, set the shift operator $\bbS$ to
\begin{align}\label{E:rankoneshif_def}
	S_{ij}=	\begin{cases}
		\mu + \sum_{n\in\ccalN_i}a_n b_n &\text{if}\;i=j,\\
		S_{ij} = -a_ib_j&\text{if}\;(i,j)\in \ccalE_{\ccalT},\\
		S_{ij}=0& \text{otherwise},
		\end{cases} 
\end{align}
where the neighborhood $\ccalN_i$ refers to the spanning tree $\ccalT$. 
We will show that the three conditions in Proposition~\ref{P:Perfect_Impl_NodeInvarFilter} are satisfied for $\bbH = \sum_{l=0}^{N-1} c_l \bbS^l$. First notice that since $D \leq N$, condition \emph{c)} is fulfilled. To see that condition \emph{a)} is satisfied, we must show that $\bbB_{\mathrm{rk1}}$ and $\bbS$ share the eigenvectors. Since $\bbB_{\mathrm{rk1}}$ is rank one, this requires showing that $\bba$ is an eigenvector of $\bbS$. To this end, define $\bbq:=\bbS\bba$. Given the construction of $\bbS$ in \eqref{E:rankoneshif_def}, it holds that $q_i = S_{ii} a_i +  \sum_{j\in\ccalN_i}S_{ij} a_j = \mu a_i$, so that $\bba$ is an eigenvector of $\bbS$ with eigenvalue $\mu$. Since the shift is not symmetric, we also need to show that the row of $\bbV^{-1}$ associated with eigenvector $\bba$ corresponds to $\bbb$. To do so, define $\bbp:=\bbS^T\bbb$ and notice that $p_j = S_{jj} b_j +  \sum_{i\in\ccalN_j}S_{ij} b_i = \mu b_j$ so that $\bbb$ is an eigenvector of $\bbS^T$ with eigenvalue $\mu$. Since $\bbS^T=(\bbV^{-1})^T\bbLambda\bbV^T$, $\bbb$ is the column of $(\bbV^{-1})^T$ (alternative the row of $\bbV^{-1}$) associated with eigenvalue $\mu$. To conclude the proof, we need to show that the multiplicity of $\mu$ is one [cf. condition \emph{b)} in Proposition~\ref{P:Perfect_Impl_NodeInvarFilter}]. 

To show that the eigenvalue $\mu$ is indeed simple, consider the incidence matrix $\bbM_{\bba}\in\reals^{N\times N-1}$ of the spanning tree. The $n$-th row of $\bbM_{\bba}$ corresponds to node $n$, while each column corresponds to one link of $\ccalT$. Suppose that the $l$-th link is given by $(i,j)$ where we follow the convention that $i<j$, then the incidence matrix $\bbM_{\bba}$ is defined as $[\bbM_{\bba}]_{il}=a_j$, $[\bbM_{\bba}]_{jl}=-a_i$ and $[\bbM_{\bba}]_{ul}=0$ for all $u\neq i$ and $u\neq j$. 
\begin{mylemma}
	If $a_i\neq 0$ for all $i$, then $\ker(\bbM_{\bba}^T)=\spanbasis\{\bba\}$ and $\ker(\bbM_{\bba})= \{ \bbzero \}$.  
\end{mylemma}
\begin{myproof} To prove that $\ker(\bbM_{\bba}^T)=\spanbasis\{\bba\}$, we first show that $\bbM_{\bba}^T\bba=\bbzero$. Using the definition of $\bbM_{\bba}$, it holds that for any $\bbx\in \reals^N$ we have $ [\bbM_{\bba}^T\bbx]_l=a_jx_i - a_ix_j$, where we recall that $(i,j)$ is the $l$-th link in $\ccalE_{\ccalT}$. Upon setting $\bbx=\bba$ it readily follows that $\bba\in \ker(\bbM_{\bba}^T)$. To show that $\bba$ is the only eigenvector with zero eigenvalue, consider matrix $\bbL_\bba \! = \! \bbM_\bba\bbM_\bba^T$. Since $[\bbM_{\bba}^T\bbx]_l=a_jx_i - a_ix_j$, it readily follows that $\bbx^T \bbL_\bba \bbx = \sum_{(i,j)\in\ccalE_{\ccalT} }(a_jx_i - a_ix_j)^2$. In order for $\bbx^T \bbL_\bba \bbx$ to be zero, all the terms $(a_jx_i - a_ix_j)^2$ must be zero. Since the graph is connected this is only possible if $\bbx=\alpha\bba$. Hence, $\rank(\bbL_\bba)=N-1$. Since $\bbL_\bba=\bbM_\bba\bbM_\bba^T$, the previous finding implies that $\rank(\bbM_\bba^T)=N-1$; hence $\ker(\bbM_{\bba}^T)$ has dimension one. Note that the fact that $\rank(\bbM_\bba^T)=N-1$ also implies that $\rank(\bbM_\bba)=N-1$, which implies that the dimension of $\ker(\bbM_{\bba})$ is zero, concluding the proof.
\end{myproof}

\begin{mylemma}
	Consider the shift $\bbS$ given in \eqref{E:rankoneshif_def} and suppose that $a_i\neq 0$ and $b_i\neq 0$ for all $i$, then it holds that $\ker(\bbS- \mu\bbI)=\spanbasis\{\bba\}$.  
\end{mylemma}
\begin{myproof}
The key step to show this result is to notice that, given \eqref{E:rankoneshif_def}, the shift can be written as $\bbS \! =\! \mu\bbI + \bbM_\bbb \bbM_\bba^T$. Using now the facts that $\ker(\bbM_{\bba}^T) \! = \! \spanbasis\{\bba\}$ (Lemma 1) and $\ker(\bbM_{\bbb}) \! = \! \{ \bbzero \}$ (Lemma 1 with $\bba \! =\! \bbb$), Lemma 2 follows. 
\end{myproof}

\noindent From Lemma 2, it follows that $\bba$ is the only eigenvector of $\bbS- \mu\bbI$ with null eigenvalue and, thus, the only eigenvector of $\bbS$ with eigenvalue equal to $\mu$, as wanted. Finally, if we combine the facts that the eigenvalue in $\bbS$ associated with $\bba$ is simple (non-repeated) and that every other eigenvalue is repeated in $\bbB_{\text{rk1}}$ (since they are equal to 0), fulfillment of condition \emph{b)} follows, completing the proof. Note that the fact of $\ccalG$ being connected is critical to guarantee that the multiplicity of $\mu$ is one. In fact it is straightforward to show that if the graph is not connected then the multiplicity of $\mu$ is equal to the number of components, so that the linear transformation cannot be implemented as a filter. 
\end{myproof}

\noindent The proof of Proposition \ref{P:shift_rankone} not only guarantees the existence of an $\bbS$ for which $\bbB_{\mathrm{rk1}} = \mathbf{a} \mathbf{b}^T$ can be written as a GF, but also provides a particular shift that achieves so [cf. \eqref{E:rankoneshif_def}]. Clearly, the definition in \eqref{E:rankoneshif_def}, which can be understood as a ``generalized'' Laplacian using the entries of $\bba$ and $\bbb$ as weights, is not unique. Note first that different spanning trees will give rise to different valid $\bbS$. Moreover, if $\bbB_{\text{rk1}}$ is symmetric, our proof can be adapted to hold for a shift $\bbS$ where the edge set in \eqref{E:rankoneshif_def} corresponds to that of any connected subgraph of $\ccalG$, {including $\ccalG$ itself. This is particularly relevant in the context of distributed implementation, since the interplay between the number of edges of the subgraph and the order of the filter can be exploited to reduce the number of local exchanges required to compute the rank-one transformation distributedly.}   

When the linear transformation $\bbB$ is of high rank, the design of a shift $\bbS$ that preserves the sparsity of $\ccalG$ and shares the eigenvectors of $\bbB$ is more involved and not always feasible. One possibility to address the design is to leverage the results in \cite{segarra2016topologyid} and obtain the shift by solving a problem of the form
	\begin{align}\label{E:general_problem_topology_id} 
	\min_{ \{\bbS, \bblambda\}} \;\; f( \bbS) \quad \text{s. to }\;\bbS=\textstyle\sum_{k =1}^N \lambda_k\bbW_k, \,\, \bbS \in \ccalS,
	\end{align}
where $\bbW_k:=\bbV\diag(\bbe_k)\bbV^{-1}$. Note that in the above problem the optimization variables are effectively the eigenvalues $\bblambda=[\lambda_1,...,\lambda_N]^T$, since the constraint $\bbS=\sum_{k =1}^N \lambda_k\bbV\diag(\bbe_k)\bbV^{-1}$ forces the columns of $\bbV$ to be the eigenvectors of $\bbS$. The objective $f$ promotes desirable network structural properties, such as sparsity or minimum-energy edge weights. The constraint set $\ccalS$ imposes requirements on the sought shift, including the elements of $\bbS$ being zero for entries $(i,j)$ corresponding to edges not present in $\ccalG$. For setups where only a few eigenvectors are known or when the equality constraint $\bbS=\textstyle\sum_{k =1}^N \lambda_k\bbV\diag(\bbe_k)\bbV^{-1}$ is relaxed, which can be useful if the optimization in \eqref{E:general_problem_topology_id} is not feasible, see \cite{segarra2016topologyid}.

%%%%%%%%%%%%%%%%%%%%%%%%%%%%%%%%%%%%%%%%%%%%
\section{Optimal design of node-variant GFs}\label{S:Impl_node_var}
%%%%%%%%%%%%%%%%%%%%%%%%%%%%%%%%%%%%%%%%%%%%

%The objective in this section is to use the node-variant GFs introduced in \eqref{E:node_variant_filter} to implement a larger class of linear network transformations. 

The objective in this section is to implement pre-specified linear transformations using the node-variant GFs introduced in \eqref{E:node_variant_filter}. Since node-invariant filters are a specific instance of \eqref{E:node_variant_filter}, the class of linear network transformations that can be implemented perfectly using node-variant GFs is larger and the approximation error (when perfect implementation is not feasible) is smaller.

More specifically, given a desired linear transformation $\bbB$ we want to design the coefficient vectors $\bbc^{(l)}$ for $l = 0, \ldots, L-1$ so that the equality
\begin{equation}\label{E:linear_transf_equal_node_variant_filter}
\bbB=\bbH_{\Nv}\big(\{\bbc_i\}_{i=1}^{N},\bbS\big)=\sum_{l=0}^{L-1}\diag (\bbc^{(l)}) \bbS^l
\end{equation}
holds. Using the same structure than that in Section \ref{S:Impl_node_invar}, we first identify the conditions under which \eqref{E:linear_transf_equal_node_variant_filter} can be solved exactly, and then analyze approximate solutions.
% in Section~\ref{Ss:perfect_impl_node_var}, while approximate solutions are analyzed in Section~\ref{Ss:approx_impl_node_var}.

%---------------------------------------------------
\subsection{Conditions for perfect implementation}\label{Ss:perfect_impl_node_var}

Defining the vectors $\bbb_i:=\bbB^T\bbe_i$ and $\tilde{\bbb}_i := \bbV^T\bbb_i$, the conditions under which the equivalence in \eqref{E:linear_transf_equal_node_variant_filter} can be achieved are given in Proposition \ref{P:Perfect_Impl_NodeVarFilter} and the subsequent corollary.
%%%%%%%%%%%%%%%%%%%%%%%%%%%%%%%
%PROPOSITION
%%%%%%%%%%%%%%%%%%%%%%%%%%%%%%%
\begin{myproposition}\label{P:Perfect_Impl_NodeVarFilter}
The linear transformation $\bbB$ can be implemented using a node-variant GF $\bbH_{\Nv}\!=\!\bbH_{\Nv}(\{\bbc_i\}_{i=1}^{N},\bbS)$ of the form given in \eqref{E:node_variant_filter} if the three following conditions hold for all $i$:\\
\noindent a) $[\tilde{\bbb}_i]_k=0$ for those $k$ such that $[\bbu_i]_k=0$.\\
\noindent b) For all $(k_1,\!k_2)$ such that $\lambda_{k_1} = \lambda_{k_2}$, it holds that $[\tilde{\bbb}_i]_{k_1}/[\bbu_i]_{k_1}=[\tilde{\bbb}_i]_{k_2}/[\bbu_i]_{k_2}$. \\
\noindent c) The degree of $\bbH_\Nv$ is such that $L\geq D$.
\end{myproposition}
\begin{myproof}
We use \eqref{E:row_H_v2_a} to write the row-wise equality between $\bbH_\Nv$ and $\bbB$
\begin{equation}\label{E:system_eq_var_filter_eq_B_v0}
\bbb_i^T=\bbh_i^T=\bbu_i^T\diag(\bbPsi\bbc_i)\bbV^{-1},
\end{equation}
for all $i$. Multiplying \eqref{E:system_eq_var_filter_eq_B_v0} from the right by $\bbV$, transposing the equality, and using the definition of $\tilde{\bbb}_i$, we get
\begin{equation}\label{E:system_eq_var_filter_eq_B}
\tilde{\bbb}_i=\diag(\bbPsi\bbc_i)\bbu_i=\diag(\bbu_i)\bbPsi\bbc_i,
\end{equation}
for all $i$. Thus, for $\bbB$ to be implementable, for every node $i$ we need to find the vector $\bbc_i$ that satisfies \eqref{E:system_eq_var_filter_eq_B}.

For each $i$, partition the set $\{1, \ldots, N\}$ into three subsets of indices $\ccalK^i_1$, $\ccalK^i_2$, and $\ccalK^i_3$ where i) $\ccalK^i_1$ contains the indices $k$ such that $[\bbu_i]_k=0$; ii) $\ccalK^i_2$ contains the indices of all the eigenvalues in $\{1, \ldots, N\} \setminus \ccalK^i_1$ that are distinct as well as one index per repeated eigenvalue $\lambda_k$, and; iii) $\ccalK^i_3 = \{1, \ldots, N\} \setminus (\ccalK^i_1 \cup \ccalK^i_2)$. Thus, separate the $N$ linear equations in \eqref{E:system_eq_var_filter_eq_B} in three groups
\begin{align}
\bbE_{\ccalK_1^i}^T \tilde{\bbb}_i = \diag(\bbE_{\ccalK_1^i}^T \bbu_i) \bbE_{\ccalK_1^i}^T \bbPsi\bbc_i, \label{E:system_eq_var_filter_eq_B_010} \\
\bbE_{\ccalK_2^i}^T \tilde{\bbb}_i = \diag( \bbE_{\ccalK_2^i}^T \bbu_i) \bbE_{\ccalK_2^i}^T \bbPsi\bbc_i, \label{E:system_eq_var_filter_eq_B_020} \\
\bbE_{\ccalK_3^i}^T \tilde{\bbb}_i = \diag(\bbE_{\ccalK_3^i}^T \bbu_i) \bbE_{\ccalK_3^i}^T \bbPsi\bbc_i.\label{E:system_eq_var_filter_eq_B_030}
\end{align}
If condition \emph{a)} holds, then \eqref{E:system_eq_var_filter_eq_B_010} is true for any $\bbc_i$ because both sides of the equality are 0. Since $\ccalK_2^i$ contains no repeated eigenvalues, the matrix $\diag( \bbE_{\ccalK_2^i}^T \bbu_i) \bbE_{\ccalK_2^i}^T \bbPsi$ in \eqref{E:system_eq_var_filter_eq_B_020} has full row-rank. Given that $| \ccalK_2^i | \leq D$, condition \emph{c)} ensures that $\bbc_i$ has at least as many elements as equations in \eqref{E:system_eq_var_filter_eq_B_020}, guaranteeing the existence of a solution. Denoting by $\bbc_i^*$ a solution of \eqref{E:system_eq_var_filter_eq_B_020}, condition \emph{b)} implies that this same vector also solves \eqref{E:system_eq_var_filter_eq_B_030}. To see why this is true, notice that for every equation in \eqref{E:system_eq_var_filter_eq_B_030} (of index $k_3$) there is one in \eqref{E:system_eq_var_filter_eq_B_020} (of index $k_2$) such that $[\bbPsi \bbc^*_i]_{k_3} = [\bbPsi \bbc^*_i]_{k_2} = [\tilde{\bbb}_i]_{k_2}/[\bbu_i]_{k_2}$, where the last equality follows from the fact that $\bbc^*_i$ solves \eqref{E:system_eq_var_filter_eq_B_020}. Imposing condition \emph{b)}, we arrive to the conclusion that $[\bbPsi \bbc^*_i]_{k_3} = [\tilde{\bbb}_i]_{k_3}/[\bbu_i]_{k_3}$ and, thus, \eqref{E:system_eq_var_filter_eq_B_030} is satisfied. The simultaneous fulfillment of \eqref{E:system_eq_var_filter_eq_B_010}, \eqref{E:system_eq_var_filter_eq_B_020}, and \eqref{E:system_eq_var_filter_eq_B_030}, implies \eqref{E:system_eq_var_filter_eq_B}, concluding the proof.
\end{myproof}
%%%END OF THE PROOF OF THE PROPOSITION%%%

The conditions in Proposition~\ref{P:Perfect_Impl_NodeVarFilter} detail how the spectral properties of $\bbS$ impact the set of linear transformations that can be implemented. Condition \emph{a)} states that if \emph{node} $i$ is unable to sense a specific frequency $k$, only linear operators whose $i$-th row is orthogonal to the $k$-th frequency basis vector can be implemented. Condition \emph{b)} states that if two frequencies $k_1$ and $k_2$ are indistinguishable for the \emph{graph-shift} operator, then the projection of the $i$-th row of $\bbB$ onto these two frequency basis vectors must be proportional to how strongly node $i$ senses frequencies $k_1$ and $k_2$ for every node $i$. Finally, condition \emph{c)} requires that the order of the \emph{filter} has to be high enough to have enough degrees of freedom to design the linear operator and to allow the original signal $\bbx$ to percolate through the network.
As was the case for Proposition~\ref{P:Perfect_Impl_NodeInvarFilter}, \emph{a)} and \emph{b)} are necessary conditions while \emph{c)} details a sufficient filter degree for general implementation. However, filters with lower degree may be enough to implement particular linear transformations.

The following result follows as a corollary of Proposition~\ref{P:Perfect_Impl_NodeVarFilter}.
%%%%%%%%%%%%%%%%%%%%%%%%%%%%%%%
%COROLLARY
%%%%%%%%%%%%%%%%%%%%%%%%%%%%%%%
\begin{mycorollary}\label{Coro:Perfect_Impl_NodeInvarFilter}
Suppose that spectrum of the graph-shift operator $\bbS = \bbV \bbLambda \bbV^{-1}$ satisfies the following two properties:\\
\noindent a) all the entries of $\bbV$ are non-zero.\\
\noindent b) all the eigenvalues $\{\lambda_k\}_{k=1}^N$ are distinct.\\
Then, it holds that any linear transformation $\bbB$ can be implemented using a node-variant GF  $\bbH_{\Nv}=\bbH_{\Nv}(\{\bbc_i\}_{i=1}^N,\bbS)$ of the form given in \eqref{E:node_variant_filter} for $L=N$, with the set of filter coefficients $\{\bbc_i\}_{i=1}^N$ being unique and given by
\begin{equation}\label{E:unique_set_filter_coefficients}
\bbc_i = \bbPsi^{-1} \diag(\bbu_i)^{-1} \bbV^T \bbb_i,\;\;\forall \;\;i.
\end{equation} 

\end{mycorollary}
\begin{myproof}
Conditions \emph{a)} and \emph{b)} immediately guarantee the corresponding conditions in Proposition~\ref{P:Perfect_Impl_NodeVarFilter} for any $\bbB$. Condition \emph{c)} in Proposition~\ref{P:Perfect_Impl_NodeVarFilter} is satisfied from the fact that the corollary sets $L = N$ and, by definition, $N \geq D$. Equation \eqref{E:unique_set_filter_coefficients} follows directly after substituting $\tbb_i=\bbV^T\bbb_i$ into  \eqref{E:system_eq_var_filter_eq_B}. Since conditions a) and b) guarantee that the full rank matrices $\diag(\bbu_i)^{-1}$ and $\bbPsi^{-1}$ exist, respectively, uniqueness also follows.  
\end{myproof}
%END OF THE COROLLARY%%%%%%%%%%%

%\label{E:row_H_v2_a}
To gain some intuition on the design in \eqref{E:unique_set_filter_coefficients}, recall first that, as explained after \eqref{E:row_H_v2_a}, the output of a node-variant GF at node $i$ (here $\bbb_i^T\bbx$) can be viewed as the inner product between $\bbu_i$ and $\widehat{\bbx}$ modulated by the frequency response of the GF $\widehat{\bbc}_i=\bbPsi\bbc_i$. Then, comparing the expression in \eqref{E:unique_set_filter_coefficients} to its counterpart for node-invariant GFs in \eqref{E:syst_equations_to_solve_perfect_case}, we observe that the design still tries to match the frequency response of $\bbB$ and $\bbH_{\Nv}$, but here in a per-node fashion; so that \eqref{E:unique_set_filter_coefficients} replaces $\bbbeta$ (the spectrum of $\bbB$) with $\bbV^T\bbb_i$ and accounts for $\bbu_i$ (the effect of the particular node in the frequency domain, which was not present in \eqref{E:syst_equations_to_solve_perfect_case}). More importantly, Proposition~\ref{P:Perfect_Impl_NodeVarFilter} and Corollary~\ref{Coro:Perfect_Impl_NodeInvarFilter} confirm that the class of linear transformations that can be implemented using \eqref{E:node_variant_filter} is quite broad and significantly larger than the one that can be implemented by using \eqref{E:def_graph_filters}. This is consistent with the fact that, as explained in Section \ref{Ss:nodevariant_filters}, node-variant GFs are more flexible operators since their number of free parameters (coefficients) is $N$ times larger than those for node-invariant filters.

%---------------------------------------------------
\subsection{Approximate implementation}\label{Ss:approx_impl_node_var}

When the conditions in Proposition~\ref{P:Perfect_Impl_NodeVarFilter} are not satisfied, we resort again to approximate designs aimed at minimizing a pre-specified error metric. As done in Section~\ref{Ss:approx_impl_node_invar} we minimize the MSE and the WCE under the assumption that the signal $\bbx$ is zero-mean and that its covariance $\bbR_\bbx$ is known. To be specific, recall that the difference (error) vector is defined as $\bbd = \bbH_\Nv \bbx - \bbB \bbx$ and its covariance matrix is denoted by $\bbR_\bbd$. Recall also that $\bbC = [\bbc_1, \ldots , \bbc_N]$ and define the matrices $\tilde{\bbU} := [\diag(\bbu_1), \diag(\bbu_2), \ldots, \diag(\bbu_N)]^T$ and $\bbPhi_i := (\bbV^{-1})^T \diag(\bbu_i) \bbPsi$. Then, the filter coefficients minimizing the MSE given by $\Tr(\bbR_\bbd)$ and the WCE given by $\|\bbR_\bbd\|_2$ can be found as specified in the following proposition, which reveals that the optimal MSE coefficients can be designed separately across nodes.

%
%PROPOSITION
\begin{myproposition}\label{P:Imperfect_Impl_NodevarFilter_2}
The optimal MSE coefficients $\{\bbc_{i, \mathrm{Tr}}^*\}_{i=1}^N \!\! := \!\! \argmin_{\{\bbc_i\}_{i=1}^N} \Tr(\bbR_{\bbd})$ are given by
\begin{equation}\label{E:optimal_filter_coefficients_case_2_var}
\bbc_{i, \mathrm{Tr}}^* = (\bbR_\bbx^{1/2} \bbPhi_i)^\dag \bbR_\bbx^{1/2} \bbb_i \stackrel{\star}{=} (\bbPhi_i^T \bbR_\bbx \bbPhi_i)^{-1} \bbPhi_i^T \bbR_\bbx \bbb_i
\end{equation}
for all $i$, where $\stackrel{\star}{=}$ holds if $\bbR_\bbx^{1/2} \bbPhi_i$ has full column rank; while the optimal WCE filter coefficients $\{\bbc_{i, \lambda}^*\}_{i=1}^N := \argmin_{\{\bbc_i\}_{i=1}^N} \lambda_{\max}(\bbR_{\bbd})$  are found as
\begin{align}\label{E:optimal_filter_coefficients_case_2_2_var}
&\{ \bbC_\lambda^*, s^*\} \,\, = \,\, \argmin_{\{\bbC, s\}} \qquad s \\
&\mathrm{s.\,to} \,\begin{bmatrix}
\! s \bbI & \!\!\!\!\! (\bbI \odot \bbPsi \bbC)^T \tilde{\bbU} \bbV^{-1} \!- \bbB \\
\! ((\bbI \odot \bbPsi \bbC)^T \tilde{\bbU} \bbV^{-1} \!- \bbB)^T & \!\!\!\!\! s \bbR_\bbx^{-1}
\end{bmatrix}
\!\!\succeq 0, \nonumber
\end{align}
where $\odot$ denotes the Khatri-Rao product.
\end{myproposition}
%%%%%%%%
\begin{myproof}
Leveraging equivalence \eqref{E:proof_Imperfect_Impl_NodeInvarFilter_2_010}, we may rewrite
\begin{equation}\label{E:trace_sum_norm_2_node_var}
\Tr(\bbR_{\bbd}) = \sum_{i=1}^N \| \bbh_i^T \bbR_\bbx^{1/2} - \bbb_i^T \bbR_\bbx^{1/2} \|_2^2.
\end{equation}
From \eqref{E:row_H_v2_a}, we write $\bbh_i$ in terms of $\bbc_i$ as $\bbh_i^T = \bbu_i^T\diag(\bbPsi\bbc_i)\bbV^{-1}$. By transposing both sides of the equation and recalling that for general vectors $\diag(\bba) \bbb = \diag(\bbb) \bba$, it follows that $\bbh_i = (\bbV^{-1})^T\diag(\bbu_i) \bbPsi\bbc_i$. Using the definition of $\bbPhi_i$ we obtain
\begin{equation}\label{E:trace_sum_norm_2_node_var_020}
\Tr(\bbR_{\bbd}) = \sum_{i=1}^N \| \bbR_\bbx^{1/2} \bbPhi_i \bbc_i - \bbR_\bbx^{1/2} \bbb_i \|_2^2.
\end{equation}
Since the $i$-th term in the summation depends on $\bbc_i$ and not on $\bbc_j$ for $i \neq j$, we may minimize the summands separately. The result in \eqref{E:optimal_filter_coefficients_case_2_var} follows from minimizing $\| \bbR_\bbx^{1/2} \bbPhi_i \bbc_i - \bbR_\bbx^{1/2} \bbb_i \|_2^2$ via the Moore-Penrose pseudoinverse \cite{Golub_MatComp_Book}.

The proof to show \eqref{E:optimal_filter_coefficients_case_2_2_var} is analogous to that of Proposition~\ref{P:Imperfect_Impl_NodeInvarFilter_2}, thus, we only need to show that $(\bbI \odot \bbPsi \bbC)^T \tilde{\bbU} \bbV^{-1} = \bbH_\Nv$. To see why this is true, transpose \eqref{E:row_H_v2_a} to obtain $\bbh_i = (\bbV^{-1})^T \diag(\bbu_i)\bbPsi\bbc_i$, from where it follows that
\begin{equation}\label{E:proof_Imperfect_Impl_NodevarFilter_1_2_010}
\![\bbh_1, ..., \bbh_N] \! = \! (\bbV^{-1})^{ \!T}  [\diag(\!\bbu_1\!), ..., \diag(\!\bbu_N\!)] \!\!
\begin{bmatrix} \bbPsi \bbc_1  \!& \!\!\! \mathbf{0}   \!& \!\!\!\cdots \\
\! \mathbf{0}  \!& \!\!\! \bbPsi \bbc_2  \!&  \!\!\!\!\\
\! \vdots  \!& \!\!\! \vdots  \! & \!\!\!\ddots
\end{bmatrix}
\!\!.\nonumber
\end{equation}
Noting that the rightmost matrix can be written as $\bbI \odot \bbPsi \bbC$ and that $\bbH_\Nv = [\bbh_1, ... , \bbh_N]^T$, the result follows.
\end{myproof}
%%%END OF THE PROOF OF THE PROPOSITION%%%

The reason that allows optimizing the MSE coefficients of a given node separately from those of the other nodes is that the MSE is given by $\Tr(\bbR_\bbd) = \| \bbH_\Nv \bbR^{1/2}_\bbx - \bbB \bbR^{1/2}_\bbx \|_\mathrm{F}^2$ [cf. \eqref{E:proof_Imperfect_Impl_NodeInvarFilter_2_010}], which is an element-wise norm that can be decoupled across nodes. By contrast, if the objective is to optimize the WCE given by $\| \bbR_{\bbd} \|_2$, then the WCE coefficients must be optimized jointly.

Naturally, there is a resemblance between the optimal designs in Proposition~\ref{P:Imperfect_Impl_NodevarFilter_2} and those for node-invariant GFs in Proposition~\ref{P:Imperfect_Impl_NodeInvarFilter_2}. When comparing \eqref{E:optimal_filter_coefficients_case_2_2} with \eqref{E:optimal_filter_coefficients_case_2_2_var}, we see that the expression $\bbV \diag(\bbPsi \bbc) \bbV^{-1}$ that explicitly states the dependence of a node-invariant filter on the filter coefficients $\bbc$, is replaced by the more involved expression $(\bbI \odot \bbPsi \bbC)^T \tilde{\bbU} \bbV^{-1}$ stating the dependence of node-variant filters on the matrix $\bbC$, which collects the filter coefficients at every node {[cf. \eqref{E:row_H_v2_a}]}. Regarding expressions \eqref{E:optimal_filter_coefficients_case_2} and \eqref{E:optimal_filter_coefficients_case_2_var}, their main difference stems from the additional design flexibility of node-variant GFs. To be more precise, if  \eqref{E:trace_sum_norm_2_node_var_020} is augmented with $\bbc_i = \bbc_j$ for all nodes $i$ and $j$, then the optimal coefficients boil down to those in \eqref{E:optimal_filter_coefficients_case_2}. The fact that every node $i$ can select different coefficients $\bbc_i$ decouples the cost in \eqref{E:trace_sum_norm_2_node_var_020} ultimately leading to the optimal expression in \eqref{E:optimal_filter_coefficients_case_2_var}.

As in Section~\ref{Ss:approx_impl_node_invar}, when no prior information about $\bbx$ is available, the covariance matrix of the input in \eqref{E:optimal_filter_coefficients_case_2_var} and \eqref{E:optimal_filter_coefficients_case_2_2_var} is set to $\bbR_{\bbx}=\bbI$. The resultant coefficients $\{\bbc_{i,\mathrm{Tr}}^*\}_{i=1}^N$ minimize the error metric $\| \bbH_\Nv - \bbB \|_{\mathrm{F}}$, while $\{\bbc_{i,\lambda}^*\}_{i=1}^N$ are the minimizers of $\| \bbH_\Nv - \bbB \|_{2}$.

Although counterparts to the results in Section \ref{Ss:Designing_the_shift} could be derived here too, optimization of $\bbS$ in this case is not as relevant, since the role of $\bbS$ in the class of transformations that node-variant GFs can implement is marginal compared to that for node-invariant GFs (see Propositions \ref{P:Perfect_Impl_NodeInvarFilter} and \ref{P:Perfect_Impl_NodeVarFilter}).
%The ensuing two sections illustrate the perfect and approximate implementations of linear network operators using node-invariant and node-variant GFs.

%%%%%%%%%%%%%%%%%%%%%%%%%%%%%%%%%%%%%%%%%%
\section{Implementation of distributed operators}\label{S:distributed_network_operators}
%%%%%%%%%%%%%%%%%%%%%%%%%%%%%%%%%%%%%%%%%%

{While the findings in Sections \ref{S:Impl_node_invar} and \ref{S:Impl_node_var} hold for any linear operator $\bbB$, our results, especially those related to node-variant GFs, are particularly relevant in the context of \textit{distributed signal processing}. To demonstrate this, we specialize $\bbB$ to {two operators} commonly studied in \textit{networked} setups:} finite-time average consensus and analog network coding (ANC). Since the leniency of the consensus operator facilitates its implementation, the focus there is primarily on node-invariant GFs. As shown next, the more challenging ANC operators call for the implementation of node-variant GFs.

%---------------------------------------------------
\subsection{Finite-time average consensus}\label{S:finite_time_consensus}

Define the $N\times N$ matrix $\bbB_{\mathrm{con}} := \mathbf{1}\mathbf{1}^T/N$ and consider the case where the goal is to implement $\bbB=\bbB_{\mathrm{con}}$. Clearly, the application of this transformation to a signal $\bbx$ yields the \textit{ consensus} signal $\bbB_{\mathrm{con}} \bbx=\mathbf{1}(\mathbf{1}^T\bbx)/N=\bbone \bar{x}$, i.e., a constant signal whose value is equal to $\bar{x}=N^{-1}\sum_{i=1}^{N}x_i$. The fact of $\bbB_{\mathrm{con}}$ being a rank-one matrix can be leveraged to show the following corollary of Proposition~\ref{P:shift_rankone}.

%%%%%%%%%% COROLLARY
\begin{mycorollary}\label{C:consensus_perfect}
The consensus transformation $\bbB_{\mathrm{con}}$ can be written as a (node-invariant) filter $\sum_{l=0}^{N-1} c_l \bbS^l$ for some $\bbS$ associated with an undirected graph $\ccalG$ if and only if $\ccalG$ is connected.
\end{mycorollary}
%%%%%%%
%\begin{myproof}
%If $\ccalG$ is disconnected such that nodes $i$ and $j$ belong to different connected components, then $\bbS_{ij}^l = 0$ for all $l=0, \ldots, N-1$ and all possible $\bbS$. This implies that matrix $\bbB_{\mathrm{con}}$, which is irreducible, cannot be written as a polynomial of any $\bbS$. Conversely, if $\ccalG$ is connected, set, for example, $\bbS = \bbI - \bbL$ where $\bbL$ is the Laplacian of $\ccalG$. We will show that, for this $\bbS$, the three conditions in Proposition~\ref{P:Perfect_Impl_NodeInvarFilter} are satisfied. First notice that since $D \leq N$, condition \emph{c)} is fulfilled. To see that condition \emph{a)} is satisfied, note that $\bbB_{\mathrm{con}}$ and $\bbS$ share the eigenvector $\mathbf{1}/\sqrt{N}$. Moreover, the rank-1 condition of the symmetric matrix $\bbB_{\mathrm{con}}$ implies that it can be diagonalized by \emph{any} orthonormal basis containing the eigenvector $\mathbf{1}/\sqrt{N}$, in particular, it is diagonalized by the eigenbasis of $\bbS$. Finally, if we combine the facts that the eigenvalue in $\bbS$ associated to $\mathbf{1}/\sqrt{N}$ is simple (non-repeated) and that every other eigenvalue is repeated in $\bbB_{\mathrm{con}}$ (since they are equal to 0), fulfillment of condition \emph{b)} follows, completing the proof.
%\end{myproof}
%%%END OF THE PROOF OF THE PROPOSITION%%%
Although $\bbB_{\mathrm{con}}$ is a particular case of the rank-one operators considered in Proposition~\ref{P:shift_rankone}, the result in Corollary~\ref{C:consensus_perfect} is stronger. The reason for this is that $\bbB_{\mathrm{con}}$ is symmetric. As explained after Proposition~\ref{P:shift_rankone}, for symmetric transformations the proof can be modified to hold not only for spanning trees but for any connected subgraph. When the design for the shift in \eqref{E:rankoneshif_def} is particularized to $\bbB_{\mathrm{con}}$, the obtained $\bbS$ is a biased and scaled version of the Laplacian. To be more precise, if we set $\mu=1$ and chose as subgraph $\ccalT$ the graph $\ccalG$ itself, then the obtained shift is $\bbS=(\bbI+\bbL)/N$, but any other selection for $\mu$ and $\ccalT$ could also be used. Notice that an immediate consequence of Corollary~\ref{C:consensus_perfect} is that \textit{if the shift can be selected}, consensus can be achieved in finite time for every connected undirected graph \cite{finiteconsensusKibangou11, sandryhaila2014finite}, with $N-1$ being an upper bound on the number of local interactions needed for convergence. Compared to classical consensus algorithms that require infinite number of iterations, the price to pay here is that the values of $\{\lambda_k\}_{k=1}^N$ need to be known in order to compute $\bbPsi$ in \eqref{E:syst_equations_to_solve_perfect_case}. If the selected $\bbS$ has \textit{repeated eigenvalues}, then the required filter degree (number of exchanges) will be lower than $N-1$ (cf. Proposition~\ref{P:Perfect_Impl_NodeInvarFilter}.\emph{c}). For setups where the upper bound on the number of allowed exchanges is less than $D-1$, the designs presented in Sections \ref{Ss:approx_impl_node_invar} and \ref{Ss:approx_impl_node_var} can be leveraged to minimize the approximation error. Another alternative is to design an $\bbS$ that, while respecting the support of $\ccalG$, minimizes the number of required exchanges. Algorithms to carry out such an optimization can be obtained along the lines of \eqref{E:general_problem_topology_id}, but their development is left as future work. 

{In setups where $\bbS$ cannot be selected, node-variant filters are a better alternative to} approximate the consensus operator. Indeed, Corollary \ref{Coro:Perfect_Impl_NodeInvarFilter} guarantees error-free implementation for a broad class of shifts. Equally interesting, when $\bbS=\bbI+a\bbL$ with $a\neq 0$, it is not difficult to show that: i) for $L\leq D$ {node-variant GFs outperform their node-invariant counterparts} (numerical simulations show that the gain is moderate); and ii) for $L > D$ both types of filters perform equally (zero error) and, in fact, yield the same set of coefficients.

%Before presenting the numerical results, one last point worth remarking is that, in Corollary~\ref{C:consensus_perfect}, the leniency of the sufficient requirement on $\ccalG$ for perfect approximation (connectedness) is a consequence of the low-rank of $\bbB_{\mathrm{con}}$. This hints that symmetric rank-1 linear operators of the form $\bbB=\bbb\bbb^T$, which have $N-1$ repeated eigenvalues, are well-suited for (node-invariant) GF approximations. In fact, it can be shown that if $\ccalG$ is connected there always exists an $\bbS$ that has $\bbb$ as eigenvector. However, for general linear operators, as the rank increases, finding shifts $\bbS$ that preserve the local connectivity of the network while simultaneously sharing the eigenvectors of $\bbB$ (cf. Proposition~\ref{P:Perfect_Impl_NodeInvarFilter}.\emph{a}) becomes less likely.

%---------------------------------------------------
\subsection{Analog network coding}\label{S:analog_network_coding}

In the context of multi-hop communication networks, network coding is a scheme where routing nodes, instead of simply relaying the received information, combine the packets (symbols) received from different sources to perform a single transmission. The general goal is to optimize the network flow while providing error-free communication \cite{ho2008network}. Even though network coding was originally conceived for transmission of digital data in the form of packets \cite{Ahlswede00Network, LinNetCoding2003}, extensions to the transmission of analog signals have been developed under the name of ANC \cite{Kattietal07ANC, Zhangetal06ANC}. In this section, we show how GFs can be leveraged to design ANC protocols. Apart from traditional communication networks, the results presented here are also relevant for setups where there exists an inherent diffusion dynamics that percolates the information across the network as, for example, in the case of molecular and nano communication networks \cite{Nakano_etal12, Kuran201086}.

%%%%%%%%%%   F   I   G   U   R    E
\begin{figure}[t]
	\centering
	\input{figures/illustration_anc_agm.tex}
	\caption{Example of desired operators $\bbB_{\mathrm{anc}}$, $\bbB_{\ccalR}$ and $\bbB_{\ccalS \ccalR}$ in the context of ANC. A graph with $N=10$ nodes is considered. The sources are $n=3$, whose destinations are $\{1,4,6,7,10\}$, and $n=6$, whose destinations are $\{2,3,5,8,9\}$, so that every node in graph is a sink and thus $\ccalR=\ccalN$. The sources and destinations are identified in the top-left and top-right graphs, respectively. The corresponding full and reduced ANC matrices are provided at the bottom of the figure.}
	\label{fig:example_anc_theory}
\end{figure}
%%%%%%%%%%%%%%%%%%%%%

{To account for the particularities of ANC, the optimal GF-design framework presented in the previous sections has to be slightly modified. Up to this point}, we have considered $\bbB \in \reals^{N \times N}$ to be a desired transformation encompassing the \textit{whole} set of nodes. However, in ANC we are interested in the transmission of information from sources to sinks that are, in general, a \textit{subset} of the nodes in a graph. To be precise, denote by $\ccalS:=\{s_1, s_2, \ldots, s_S\}$ the set of $S$ sources and by $\ccalR:=\{r_1, r_2, \ldots, r_R\}$ the set of $R$ sinks or receivers. Since every source can have one or more receivers, we also define the (surjective) function $s:\; \ccalR \rightarrow \ccalS$, which identifies the source for each receiver. In ANC one is interested in transformations $\bbB=\bbB_{\mathrm{anc}}$ where the $i$-th row of $\bbB_{\mathrm{anc}}$ is equal to the canonical vector $\bbe_{s(i)}^T$ for all $i \in \ccalR$ (see Fig. \ref{fig:example_anc_theory} for an example). Since the values of $\bbB_{\mathrm{anc}}$ for rows $i\notin \ccalR$ are not relevant for the performance of ANC, one can define the reduced $R \times N$ matrix $\bbB_{\ccalR}:=\bbE_{\ccalR}^T\bbB_{\mathrm{anc}}$, where $\bbE_{\ccalR} := [\bbe_{r_1}, \ldots, \bbe_{r_R}]$. Hence, the goal of ANC boils down to designing a filter $\bbH$ such that $\bbE_{\ccalR}^T\bbH$ is as close to $\bbB_{\ccalR}$ as possible. Most ANC setups consider that only source nodes have signals to be transmitted, so that the input signal at all other nodes can be considered zero. The GF design can leverage this fact to yield a better approximation. Indeed, it is easy to see that if the input signal for nodes $i\notin \ccalS$ is zero, the values of the corresponding columns in $\bbB_{\ccalR}$ are irrelevant. Hence, upon defining matrices $\bbE_{\ccalS} := [\bbe_{s_1}, \ldots, \bbe_{s_S}]$ and $\bbB_{\ccalS \ccalR}:=\bbB_{\ccalR}\bbE_{\ccalS}\in \reals^{S\times  R}$, the goal for ANC is to design a filter $\bbH$ such that $\bbE_{\ccalR}^T\bbH \bbE_{\ccalS}$ is as close to $\bbB_{\ccalS \ccalR}$ as possible. To illustrate why designing $\bbB_{\ccalS \ccalR}$ is easier, consider the example in Fig. \ref{fig:example_anc_theory}. While perfect implementation of $\bbB$ requires tuning the 100 entries of $\bbH$, implementation of $\bbB_{\ccalS \ccalR}$ requires only fixing 20 entries (those corresponding to the 3rd and 6th columns). 

% := [\bbe_{s_1}, \ldots, \bbe_{s_S}]$ $  the the routing nodes (), then the desired transmission of information from $\ccalS$ to $\ccalR$ can be written as $\bbB_{\ccalS \ccalR} \in \reals^{R \times S}$. For example, if $R=S$ and every source is paired with a target such that the signal at $s_i$ must be received by $r_i$ for all $i$, then $\bbB_{\ccalS \ccalR} = \bbI$.

Although the propositions presented throughout the paper need to be modified to accommodate the introduction of $\bbB_{ \ccalR}$ and $\bbB_{\ccalS \ccalR}$, the main results and the structure of the proofs remain the same. To be specific, consider first the case where the nodes that are not sources do not inject any input, so that the goal is to approximate the $S \times R$  matrix $\bbB_{\ccalS \ccalR}$. Then, defining the $SR \times L$ matrix $\Theta_{\ccalS\ccalR}:=[\mathrm{vec}(\bbE_\ccalR^T \bbI \bbE_\ccalS), \ldots, \mathrm{vec}(\bbE_\ccalR^T \bbS^{L-1} \bbE_\ccalS)]$, we can find the coefficients that minimize $\| \bbE_\ccalR^T \bbH \bbE_\ccalS -  \bbB_{\ccalS \ccalR} \|_\mathrm{F}$ as [cf. \eqref{E:optimal_filter_coefficients_case_2}]
\begin{equation}\label{E:gen_optimal_filter_coefficients_case_1}
\bbc_\mathrm{F}^* \!=\! \bbTheta_{\ccalS\ccalR}^\dag \mathrm{vec}(\bbB_{\ccalS \ccalR}) \!\stackrel{\star}{=}\! (\bbTheta_{\ccalS\ccalR}^T \bbTheta_{\ccalS\ccalR})^{-1} \bbTheta_{\ccalS\ccalR}^T \mathrm{vec}(\bbB_{\ccalS \ccalR}),
\end{equation}
where $\stackrel{\star}{=}$ holds if $\bbTheta_{\ccalS\ccalR}$ has full column rank.
Similarly, by defining $\bbPhi_{r_i, \ccalS}:= \bbE_\ccalS^T \bbPhi_{r_i}$ and denoting the $i$-th row of $\bbB_{\ccalS \ccalR}$ as $\bbb^T_{i, \ccalS}$, we may obtain the node-variant counterpart of \eqref{E:gen_optimal_filter_coefficients_case_1} as [cf. \eqref{E:optimal_filter_coefficients_case_2_var}]
\begin{equation}\label{E:gen_optimal_filter_coefficients_case_1_var}
\bbc_{r_i, \mathrm{F}}^* = \bbPhi_{r_i, \ccalS}^\dag \bbb_{i, \ccalS} \stackrel{\star}{=} (\bbPhi_{r_i, \ccalS}^T \bbPhi_{r_i, \ccalS})^{-1} \bbPhi_{r_i, \ccalS}^T \bbb_{i, \ccalS},
\end{equation}
for all $r_i \in \ccalR$, where $\stackrel{\star}{=}$ holds if $\bbPhi_{r_i, \ccalS}$ has full column rank. For a given filter length, the $L$ coefficients in $\bbc_{r_i, \mathrm{F}}^*$ specify the optimal weights that the sink node $r_i$ must give to the original signal and the first $L-1$ shifted versions of it to resemble as close as possible (in terms of MSE) the desired linear combination of source signals $ \bbb_{i, \ccalS}$. 

When the initial signal at the routing nodes cannot be assumed zero, the goal is to approximate the %$R \times N$
matrix $\bbB_{ \ccalR}\in \reals^{R \times N}$. In that case, the previous expressions for the optimal filter coefficients still hold true. The only required modification is to substitute $\bbB_{\ccalS \ccalR}\!=\!\bbB_{\ccalR}$ and $\bbE_\ccalS\!=\!\bbI\! \in \! \reals^{N\times N}$ into the definitions of $\bbTheta_{\ccalS\ccalR}$, $\bbPhi_{r_i, \ccalS}$ and $\bbb_{i, \ccalS}$.
Indeed, when every node acts as both a source and a sink: i) $\bbB_{\ccalR}$ is an $N \times N$ matrix equal to $\bbB_{\mathrm{anc}}$; and ii) the definitions of $\bbTheta_{\ccalS\ccalR}$ and $\bbPhi_{r_i, \ccalS}$ require setting $\bbE_\ccalS=\bbE_\ccalR=\bbI$. This readily implies that \eqref{E:gen_optimal_filter_coefficients_case_1} and \eqref{E:gen_optimal_filter_coefficients_case_1_var} reduce to their original counterparts \eqref{E:optimal_filter_coefficients_case_2} and \eqref{E:optimal_filter_coefficients_case_2_var}, respectively. 

Although not presented here, expressions analogous to those in \eqref{E:gen_optimal_filter_coefficients_case_1} and \eqref{E:gen_optimal_filter_coefficients_case_1_var} for the remaining optimal filter-design criteria can be derived too.

\section{Numerical experiments}\label{S:NumExper}

This section is structured in three parts. The first two focus on illustrating the results in Sections \ref{S:Impl_node_invar} and \ref{S:Impl_node_var} using as examples the two applications introduced in Section \ref{S:distributed_network_operators}. The last part assesses the approximation error when the eigenvectors of $\bbS$ and $\bbB$ are different, which according to  Proposition \ref{P:Perfect_Impl_NodeInvarFilter} is the critical factor for approximation performance. {The results presented here complement and confirm the preliminary findings reported in \cite{ssamar_distfilters_allerton15} and  \cite{ssamar_distfilters_icassp16} for other types of graphs.}

\subsection{Finite-time average consensus}

%%%%%%%%%%   F   I   G   U   R    E
\begin{figure*}[t]
	\centering
	
	\begin{subfigure}{.33\textwidth}
		\centering
		\includegraphics[width=1\textwidth]{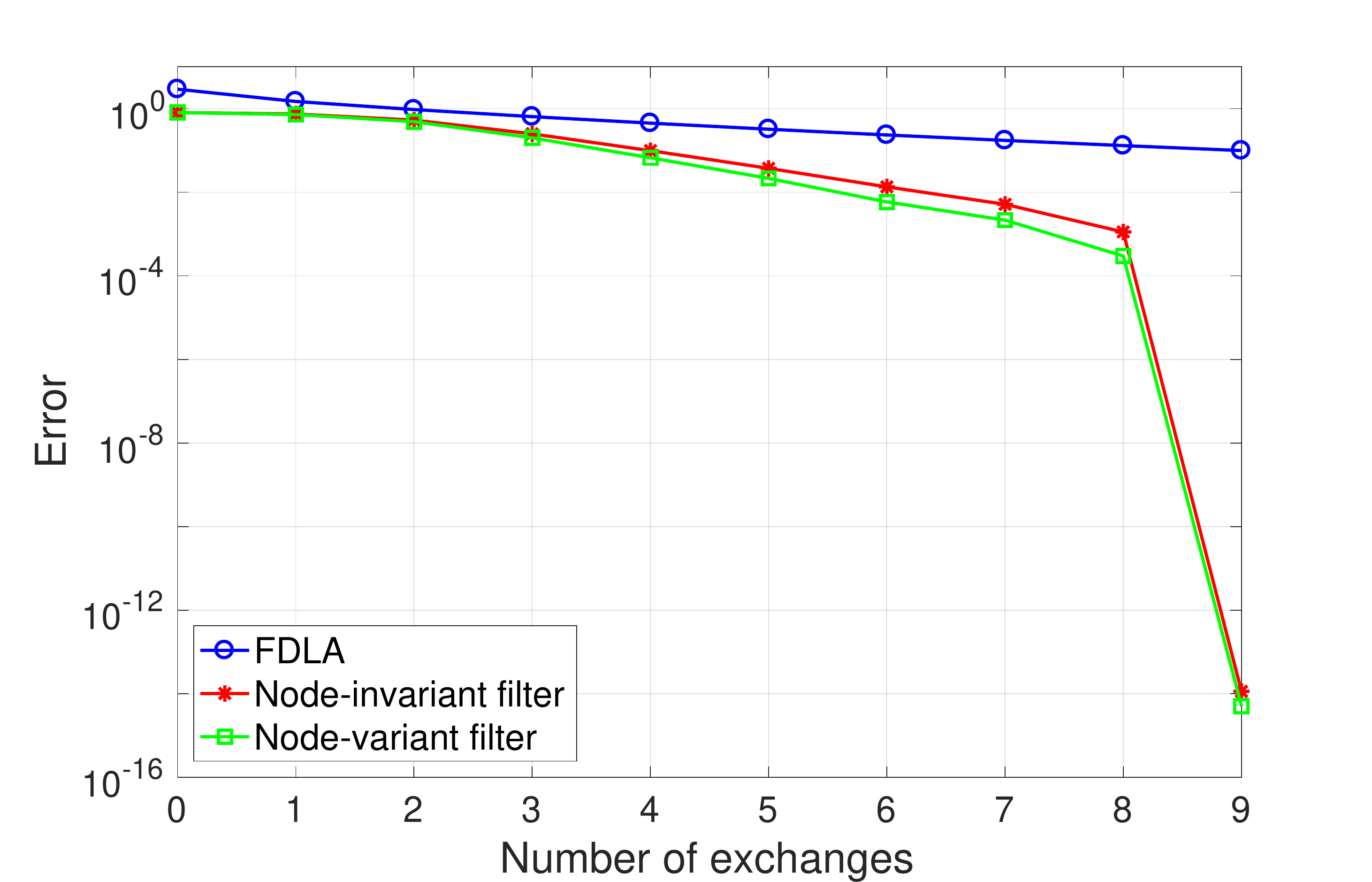}
		\caption{{Different types of filters}}
		\label{fig:sub1_consensus}
	\end{subfigure}%
	\begin{subfigure}{.33\textwidth}
		\centering
		\includegraphics[width=1\textwidth]{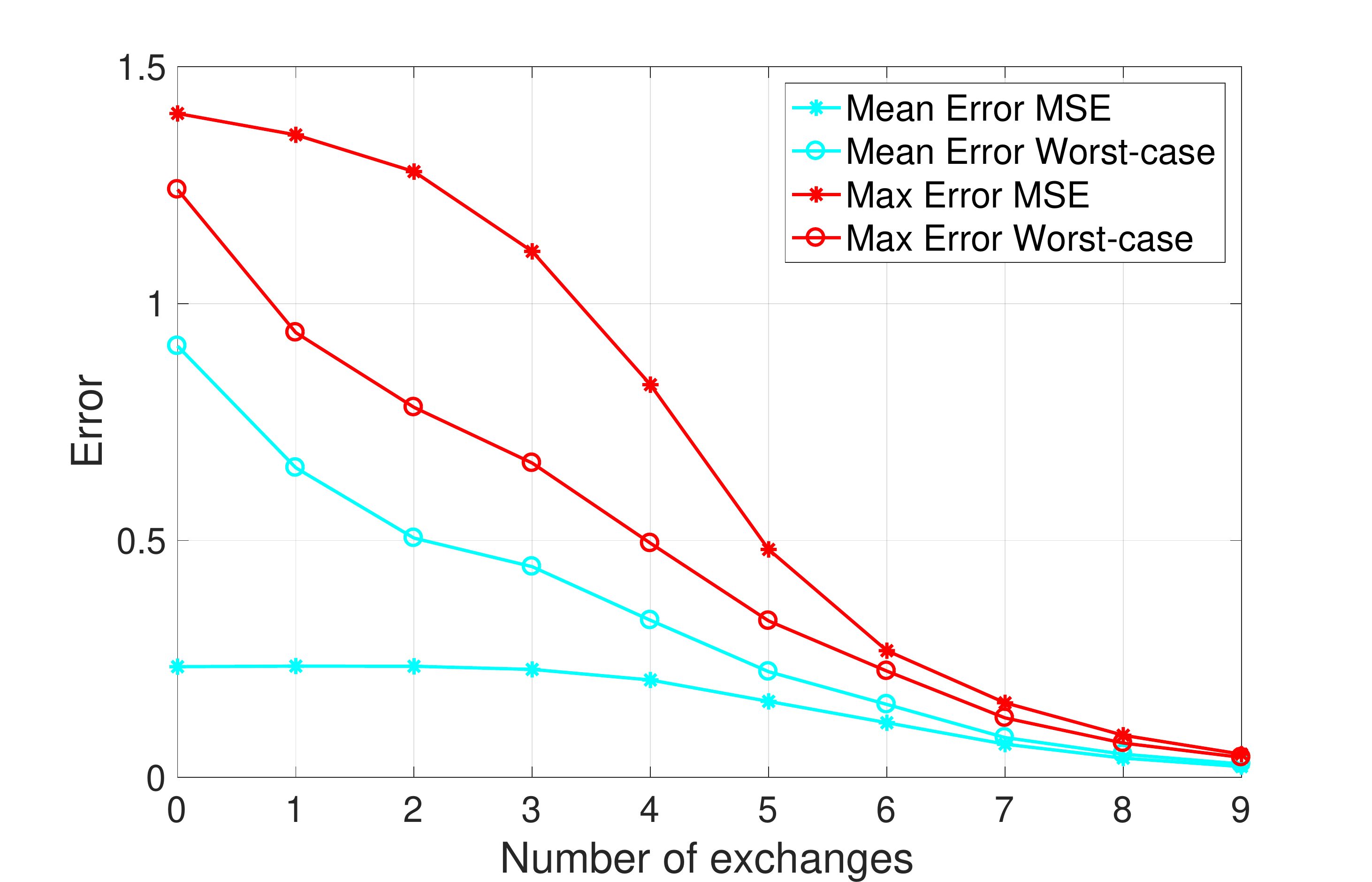}
		\caption{{Different coefficient-design criteria}}
		\label{fig:sub2_consensus}
	\end{subfigure}%
	\begin{subfigure}{.33\textwidth}
		\centering
		\includegraphics[width=1\textwidth]{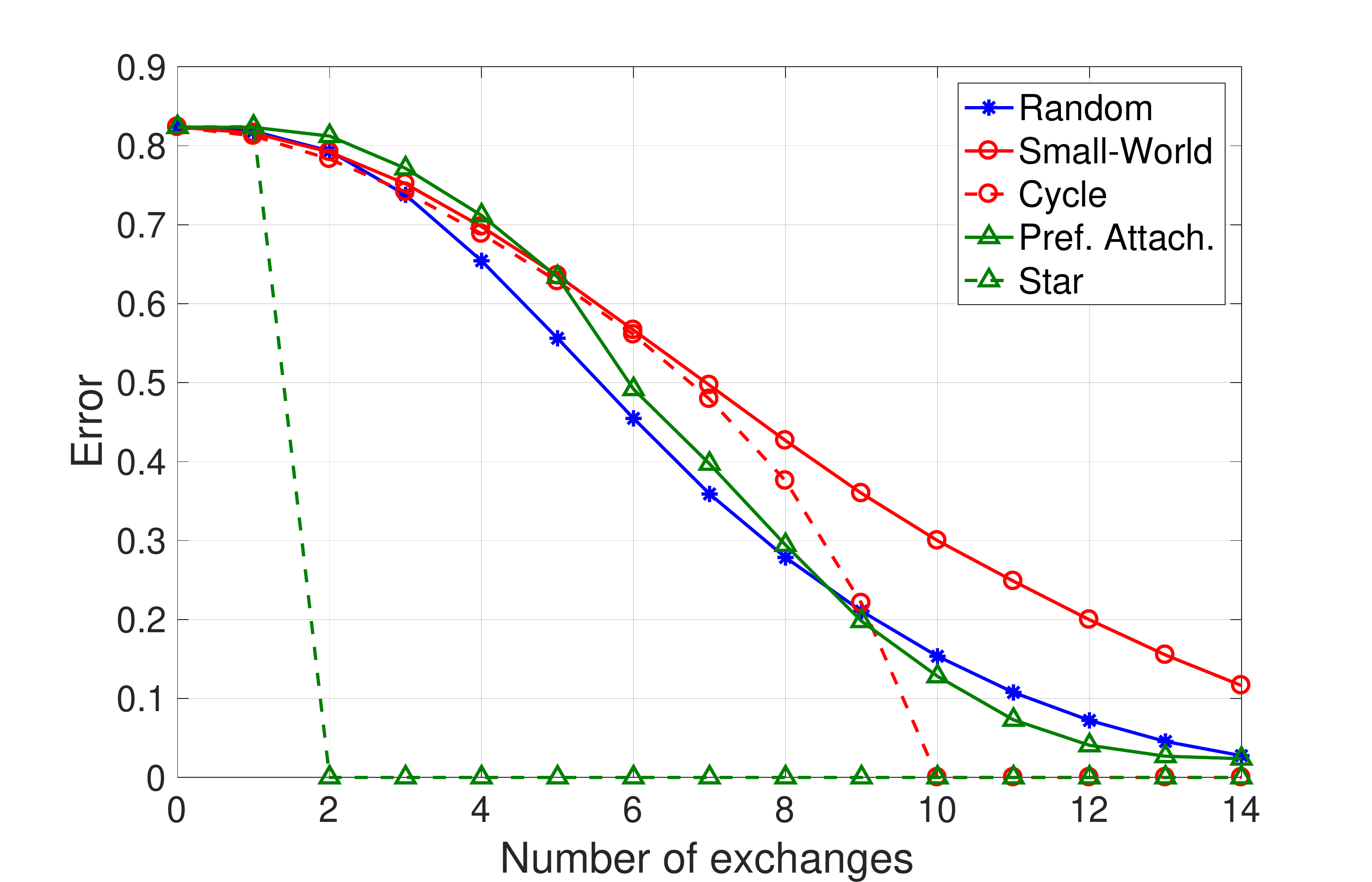}
		\caption{Different underlying graphs}
		\label{fig:sub3_consensus}
	\end{subfigure}
	\caption{(a) Mean approximation error for consensus across 1,000 realizations in 10-node {small-world} graphs. GF approaches outperform the best asymptotic solution and achieve perfect consensus for degree $N-1 = 9$. (b) Mean and maximum errors obtained across 1,000,000 signal realizations for a {40-node scale-free graph} and two different node-invariant GFs. One of the filters attains lower average error and the other attains lower maximum error, as expected. (c) Mean approximation error for consensus across 1,000 realizations of 20-node Erd\H{o}s-R\'{e}nyi, small-world, and scale-free graphs. Errors for star and cycle graphs are also shown.}
	\label{fig_consensus}
\end{figure*}
%%%%%%%%%%%%%%%%%%%%%

To demonstrate the practical interest of the proposed framework, we present numerical experiments comparing the performance achieved by three different consensus implementations: our design as a node-invariant GF, its node-variant counterpart, and the asymptotic fastest distributed linear averaging (FDLA) in \cite{XiaoBoyd2004}. In order to assess the performance of the approximation algorithms, we generate 1,000 unweighted symmetric {small-world graphs \cite{kolaczyk2009book} with $N=10$ nodes, where each node is originally connected in average to four of its neighbors, and a rewiring probability of $0.2$}. For each graph, we define the graph-shift operator $\bbS = \bbW$ where $\bbW$ is the solution of the FDLA problem, i.e, {$\lim_{l \to \infty} \bbW^l = \bbB_{\mathrm{con}}$} with fastest convergence. Moreover, on each graph we define a signal $\bbx$ drawn from a standard multivariate Gaussian distribution (smGd). For a fixed number $K$ of local interactions, we define the FDLA approximation as $\bbx_{\mathrm{FDLA}}^{(K)} := \bbW^K \bbx$, the node-invariant GF approximation as $\bbx_{\mathrm{GF}}^{(K)} := \sum_{l=0}^{K} c^*_l {\bbW}^l \bbx$ and the node-variant GF approximation as $\bbx_{\mathrm{NVGF}}^{(K)} := \sum_{l=0}^{K} \diag (\bbc^{(l)*}) {\bbW}^l \bbx$. The optimal coefficients $c^*_l$ and $\bbc^{(l)*}$, which change with $K$, are obtained for all $l$ using \eqref{E:optimal_filter_coefficients_case_2} and \eqref{E:optimal_filter_coefficients_case_2_var}, respectively. Further, we define the error $e_{\mathrm{GF}}^{(K)} = \| \bbx_{\mathrm{GF}}^{(K)} - \bbB_{\mathrm{con}}\bbx \|_2$ and similarly for $e_{\mathrm{FDLA}}^{(K)}$ and $e_{\mathrm{NVGF}}^{(K)}$. {Fig.~\ref{fig:sub1_consensus} plots} the errors averaged over the 1,000 graphs as a function of the number of local {interactions ($K \!=\! L - 1$) among neighbors}. The error attained by the GF approaches is around one order of magnitude lower than that of the asymptotic approach for intermediate number of interactions and, when $K=N-1=9$, perfect recovery is achieved using GFs (cf.~Corollary~\ref{C:consensus_perfect}). The performance improvement attained by the GFs is due to the fact that, for a fixed $K$, FDLA returns the value of $\bbW^K \bbx$ whereas the GFs return an optimal linear combination of all {$\bbW^l \bbx$ for $0 \leq l \leq K$}. Notice that node-variant GFs, being a generalization of node-invariant filters, are guaranteed to achieve a lower error. Nevertheless, the additional error reduction achieved by node-variant GFs is minimal, due to the fact that $\bbB_\mathrm{con}$ can be perfectly implemented using node-invariant filters.

We now focus {the analysis} on node-invariant filters and illustrate the difference between the MSE and the WCE minimizations (cf. Proposition~\ref{P:Imperfect_Impl_NodeInvarFilter_2}). {To this end, we generate an unweighted and symmetric scale-free graph \cite{kolaczyk2009book} with $N=40$ nodes from a preferential attachment dynamic with four initial nodes and where each new node establishes two edges with the existing graph.} On this graph, we define 1,000,000 signals $\bbx$ drawn from an smGd. For a fixed filter degree $K$, we compute the optimal MSE filter $\bbH^{(K)}_\mathrm{MSE}$ using \eqref{E:optimal_filter_coefficients_case_2}, and the optimal WCE filter $\bbH^{(K)}_\mathrm{WCE}$ using \eqref{E:optimal_filter_coefficients_case_2_2}. We then define the error $e^{(K)}_\mathrm{MSE} := ||\bbH^{(K)}_\mathrm{MSE}\bbx - \bbB_\mathrm{con} \bbx ||_2$ and similarly for $e^{(K)}_\mathrm{WCE}$. In Fig.~\ref{fig:sub2_consensus} we plot the average of these errors across the 1,000,000 realizations {as well as their maximum} as a function of the filter degree. If we focus on the average error, it is immediate to see that, as expected, the MSE approach outperforms the WCE. On the other hand, if we consider the maximum error across realizations, $\bbH^{(K)}_\mathrm{WCE}$ attains the lowest error. Notice that for $K \geq 9$, the mean and maximum errors achieved by both approaches are negligible, even though the degree $K$ is markedly smaller than $N-1=39$. 
%{These findings are consistent with those reported in \cite{ssamar_distfilters_allerton15} for Erd\H{o}s-R\'{e}nyi graphs \cite{kolaczyk2009book}.}

To consider different graph models, the last set of simulations includes {Erd\H{o}s-R\'{e}nyi \textit{random} graphs \cite{kolaczyk2009book} as well as the \textit{deterministic} star and cycle graphs. Regarding the random graphs, the} error performance is {assessed} by generating 1,000 instances of 20-node graphs for each class and setting $\bbS = \bbL$. To obtain graphs with comparable number of edges, for the Erd\H{o}s-R\'{e}nyi graphs we set $p_{edge}=0.1$, we obtain the small-world graphs by rewiring with probability $0.2$ the edges in a cycle, and we generate the scale-free graphs from a preferential attachment dynamic where each new node establishes \emph{one} edge with the existing graph. In Fig.~\ref{fig:sub3_consensus} we illustrate the mean error across 1,000 realizations as a function of the filter degree $K$. The approximation errors attained for random (Erd\H{o}s-R\'{e}nyi) and scale-free (preferential attachment) graphs are comparable, with the former showing slightly better performance for small values of $K$ while the opposite being true for large values of $K$. Moreover, the approximation errors obtained for small-world graphs are consistently larger than those for the other two types of graphs. This can be in part explained by the fact that, for the simulation setup considered, the average diameters for the Erd\H{o}s-R\'{e}nyi and the scale-free graphs are 7.0 and 6.1, respectively, whereas the average diameter for the small-world graphs is 10.9. {Fig.~\ref{fig:sub3_consensus}} also shows the consensus approximation error for {the star and the cycle, which are deterministic graphs.} Notice that the {star} {can be understood} as the limit of preferential attachment graphs, where every new node attaches to the existing node of largest degree with probability 1. Furthermore, the cycle can be seen as the limit of a small-world graph, where the probability of rewiring is 0. The star graph achieves zero error after two information exchanges between neighbors. Similarly, the cycle achieves zero error after $10 = N/2$ local interactions between neighbors. This implies that, for the star and the cycle graphs, perfect consensus is achieved for a number of exchanges equal to the diameter of the graph \cite{sandryhaila2014finite}, which is a trivial lower bound for perfect recovery since information must travel from each node to every other node for consensus to be achievable.

%As mentioned in Section~\ref{S:Introduction}, other finite-time consensus algorithms exist \cite{finiteconsensusKibangou11, sandryhaila2014finite}. However, the objective of the numerical experiments here is \emph{not} to show that the introduced method outperforms existing finite-time consensus algorithms but rather to demonstrate that it shares the benefits of finite-time algorithms while being a more general framework, as illustrated in the next sections.

\subsection{Analog network coding}

%%%%%%%%%%   F   I   G   U   R    E
%\begin{figure}[t]
%	\centering
%	\input{figures/illustration_anc.tex}
%	\caption{Example of optimal design of node-variant filters for analog network coding. Optimal estimates at sink nodes are shown by node colors for different times. After three interactions with neighbors, every sink node can decode its objective signal.}
%	\label{fig:example_anc}
%\end{figure}
%%%%%%%%%%%%%%%%%%%%%

%%%%%%%%%%   F   I   G   U   R    E
\begin{figure*}[t]
	\centering
	\hspace{0.03cm}
	\begin{subfigure}{.35\textwidth}
		\centering
		\small
		\begin{tabular}{c | c | c c c c}\small
			& recover & \multicolumn{4}{c}{observed signal $\bbz^{(t)}$}\\
			node & signal & $t=0$ & $t=1$ & $t=2$ & $t=3$ \\ \hline
			1 & $g$	& 0 	& $g$ 	& $g+w$ 	& $5g+2w$\\
			2 & $w$ 	& 0 	& $g$ 	& $g+w$	& $5g+3w$\\
			3 & $w$	&$g$	& 0 		& $4g+2w$& $4g+3w$\\
			4 & $g$	& 0	& $g + w$	& $g$	& $7g+7w$\\
			5 & $w$	& 0	& $g + w$	& $g+w$	& $5g+8w$\\
			6 & $g$	&$w$ & 0 		& $2g+4w$& $3g+2w$\\
			7 & $g$	& 0 	& $w$ 	& 0 		& $2g+5w$\\
			8 & $w$	& 0 	& $w$ 	& $g+w$	& $4g+7w$\\
			9 & $w$	& 0	& 0 		& $g+2w$	& $2g+3w$\\
			10& $g$	& 0 	& 0		& $w$	& $g+2w$\\ \hline
		\end{tabular}
		\vspace{.6cm}
		\caption{Observed signal for $t=0,..,3$}
		\label{tab:anc_example}
	\end{subfigure}%
	\hspace{2cm}
	\begin{subfigure}{.45\textwidth}
		%\centering
		%\input{figures/illustration_anc.tex}
		\hspace{.5cm}\includegraphics[width=\textwidth]{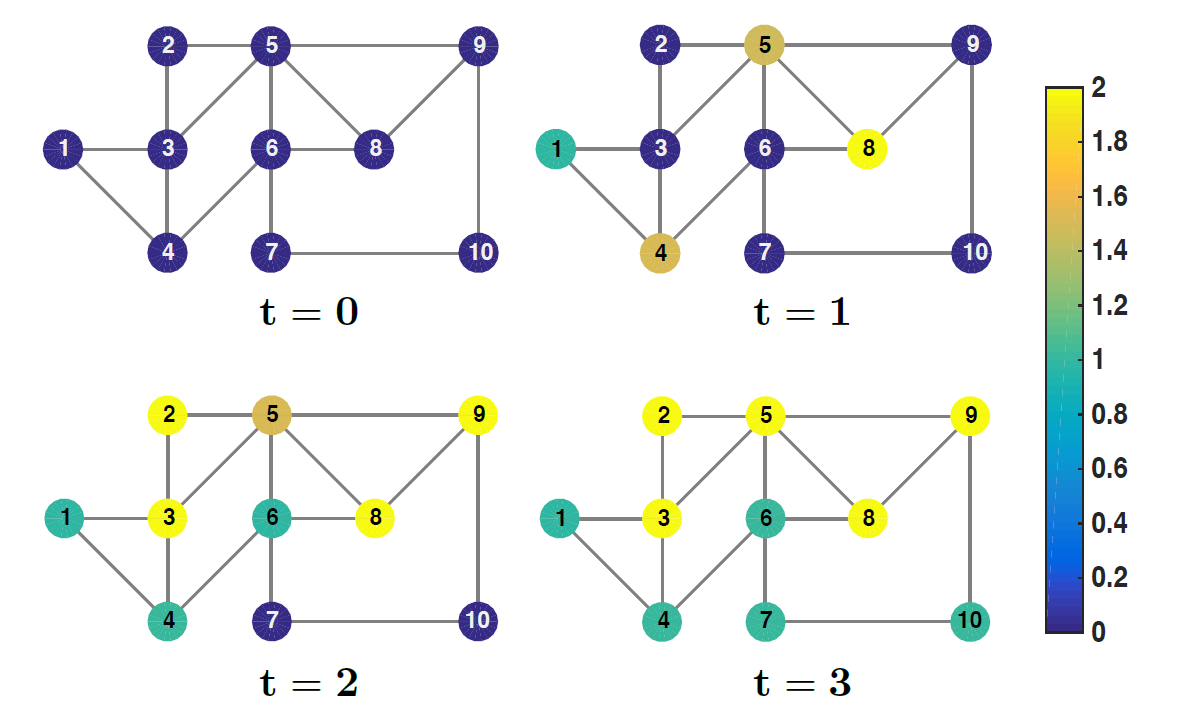}
		\caption{Best estimate of the target signal for $t=0,..,3$}
		\label{fig:example_anc}
	\end{subfigure}%

	\caption{Example of design of a node-variant GF for the ANC setup described in Fig. \ref{fig:example_anc_theory}. (a) Signal to recover at each sink node ($g$ stands for green and $w$ for yellow) and observed shifted signals for $t \in \{0, 1, 2, 3\}$. (b) Optimal estimates at sink nodes shown by node colors for different times. After three interactions with neighbors ($t=3$), the bottom-right graph shows that every sink can decode its target signal. }
	\label{figtab_ANC_combined}
\end{figure*}
%%%%%%%%%%%%%%%%%%%%%

%%%%%%%%%%%%%%%   F   I   G   U   R    E   :   R    A    N    D        N   E   T   W   O   R   K   %%%%%%%%%%%%%%%%%%%%%
\begin{figure*}[t]
	\centering	
	\begin{subfigure}{.32\textwidth}
		\centering
		\includegraphics[width=\linewidth]{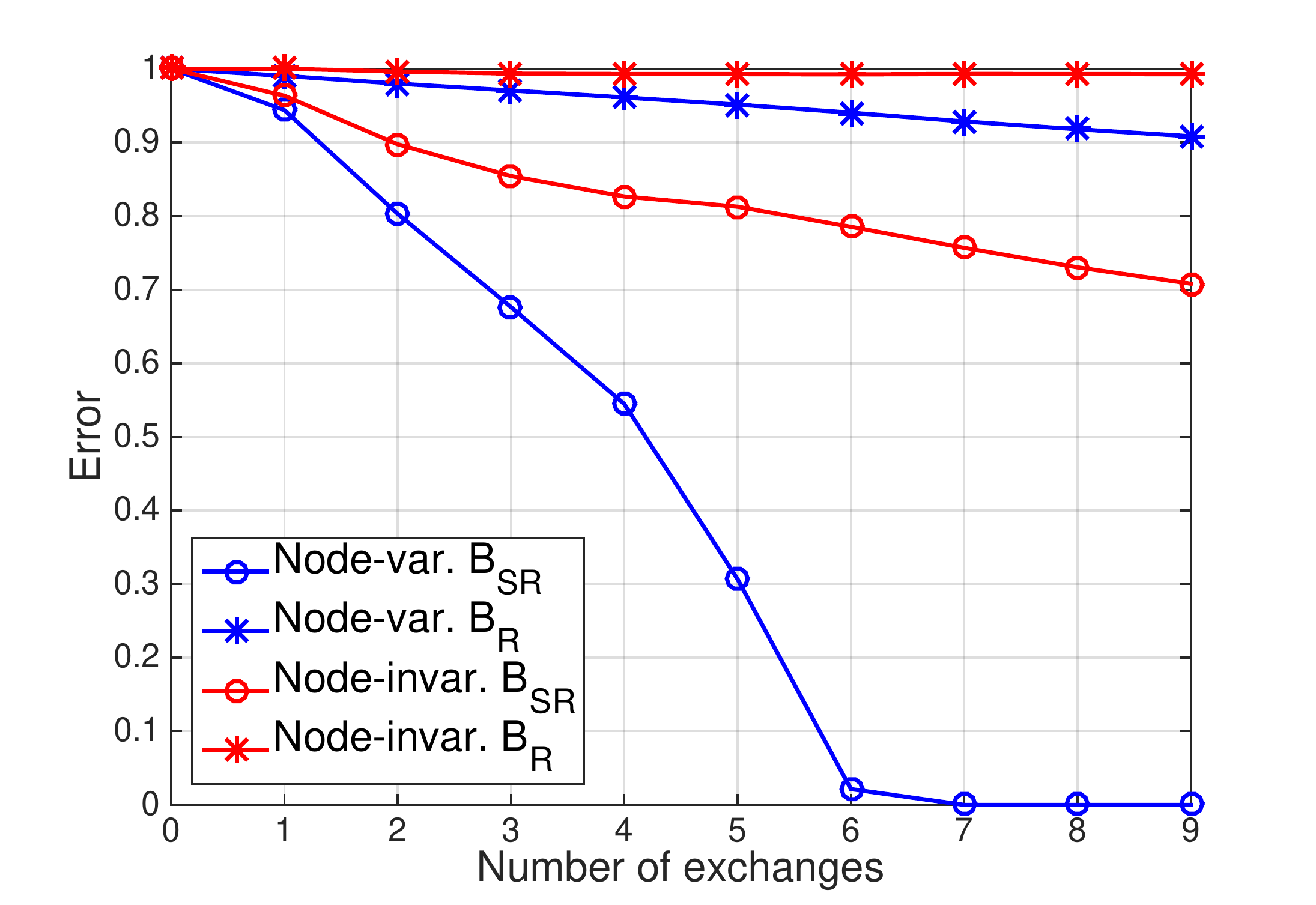}
		\caption{ANC uncorrelated sources}
		\label{fig:sub2_anc}
	\end{subfigure}
	\begin{subfigure}{.32\textwidth}
		\centering
		\includegraphics[width=\linewidth]{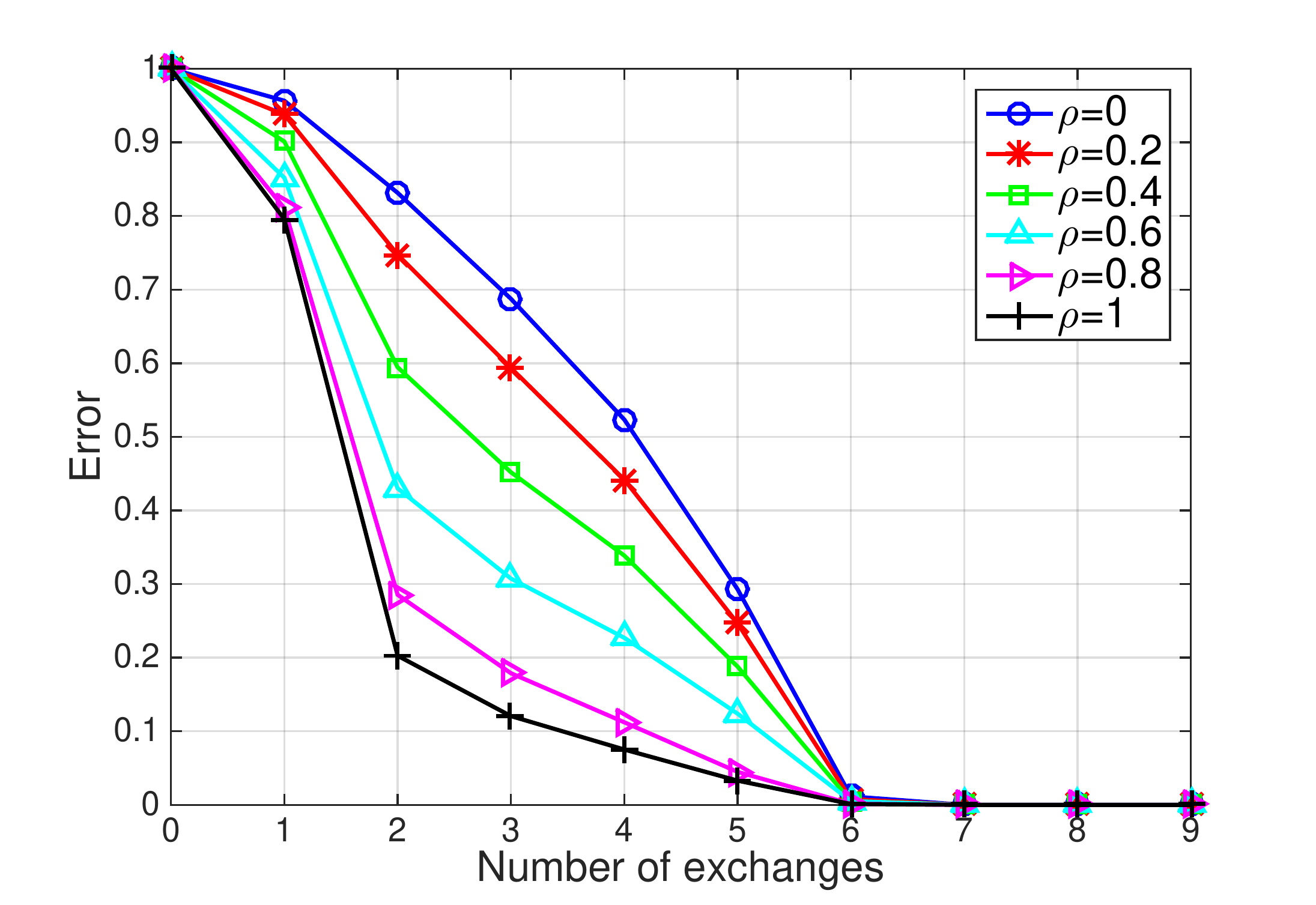}
		\caption{ANC correlated sources}
		\label{fig:sub3_anc}
	\end{subfigure}
	\begin{subfigure}{.32\textwidth}
		\centering
		\includegraphics[width=\linewidth]{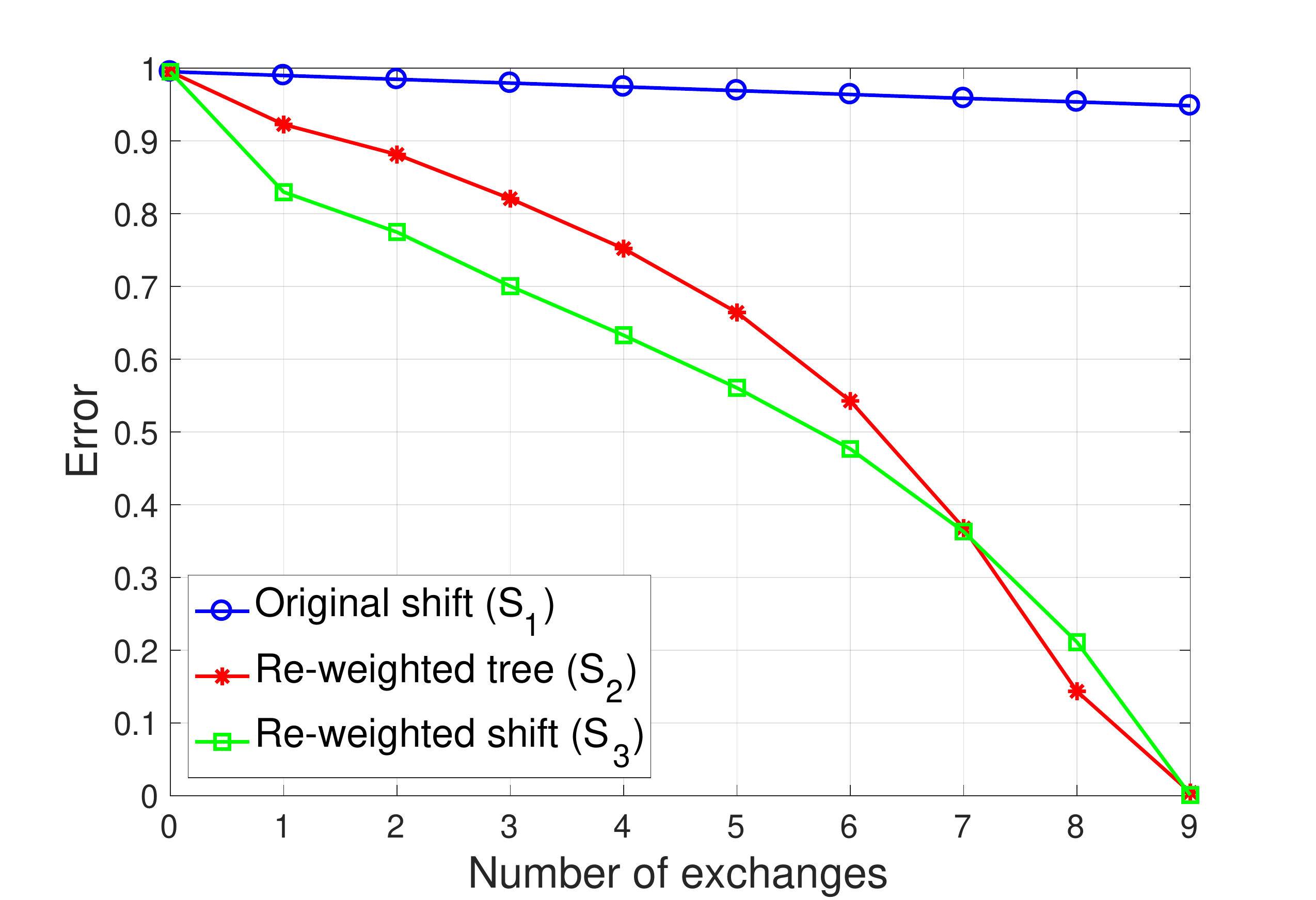}
		\caption{{Rank-one transformations}}
		\label{fig_optimal_shift_design}
	\end{subfigure}
	\caption{{Approximation errors as a function of the filter degree based on 1,000 realizations and Erd\H{os}-R\'enyi random graphs. (a) ANC with $N\!=\!100$, $S\!=\!R\!=5$ sources and sinks. Curves represent mean recovery error for node-variant and node-invariant GFs when the coefficients are designed to approximate $\bbB_{\ccalR}$ and $\bbB_{\ccalS\ccalR}$. 
		(b) Same as in (a) for node-variant GFs and different levels of correlation $\rho$ among the values injected at the different sources.
		(c) Rank-one transformations as node-invariant GFs for three different choices of graph-shift operators.}}
	\label{fig_extra_simus2}
\end{figure*}
%%%%%%%%%%%%%%%%%%%%%%%%%%%%%%%%%%%%%%%%%%%%%%%%%%%%%%%%%%%%%%%%%%%

To assess the performance obtained when node-variant GFs are designed to implement ANC schemes, we consider the example described in Fig.~\ref{fig:example_anc_theory}, where we recall that node $3$ wants to transmit its signal to nodes 1, 4, 6, 7, and 10, while node $6$ wants its signal to be transmitted to the remaining five nodes. The corresponding desired matrix $\bbB_{\ccalS\ccalR}$ is shown in the figure too. {For} this particular example, every sink node wants to recover either one source signal \emph{or} the other, but the same model could accommodate the case where sink nodes seek to recover a linear combination of the source signals. Denote by $g$ (green) the signal injected by source node 3 and by $w$  (yellow) the one injected by node 6, and set $\bbS=\bbA$. Fig.~\ref{tab:anc_example} summarizes the signal to be recovered at each node as well as the initially injected signal $\bbz^{(0)}$ and its first three shifted versions. In Fig.~\ref{fig:example_anc} we specialize this problem for the case where the green signal $g$ and the yellow signal $w$ are equal to 1 and 2, respectively. Moreover, we use the node colors to represent the successive best approximations of the signal to be recovered at each node, obtained using the optimal filter coefficients in \eqref{E:gen_optimal_filter_coefficients_case_1_var}. For example, at time $t=0$, we have that 8 of the nodes have only observed a null signal and nodes 3 and 6 have observed their own injected signal which is different from the signal they want to recover. Thus, none of the nodes has information about the signal they want to recover and the optimal estimate is 0 for all of them. However, after one shift ($t=1$) we see that four nodes have changed their estimates. The optimal coefficients $\bbc^*_{1,\mathrm{F}}=\bbc^*_{8,\mathrm{F}}=[0, 1]^T$ allow nodes 1 and 8 to perfectly recover the desired signal, as can be seen from the corresponding rows of the table in Fig.~\ref{tab:anc_example}. For example, node $1$ perfectly recovers $g$ by computing ${\bbc^{*T}_{1,\mathrm{F}}} \, [0,g]^T = g$. The optimal coefficients for nodes 4 and 5 at $t=1$ are $\bbc^*_{4,\mathrm{F}}=\bbc^*_{5,\mathrm{F}}=[0, 0.5]^T$ since their observation after one shift is the sum of the signal they want to recover and the other input signal. At time $t=2$, seven of the sink nodes perfectly recover the desired signal. E.g., node 3 applies the optimal coefficients $\bbc^*_{3,\mathrm{F}}=[-2, 0, 0.5]^T$, which yield the signal $\bbc^{*T}_{3,\mathrm{F}} \, [g,0,4g+2w]^T = w$ as can be seen from the corresponding row in Fig.~\ref{tab:anc_example}. By contrast, nodes 7 and 10 have not observed any information related to their desired signal and node 5 still cannot decode $w$ since it observed $g+w$ twice. Finally, at time $t=3$ every node can successfully decode the objective signal. 
%For example, the optimal coefficients for node 10 can be found to be $\bbc^*_{10,\mathrm{F}}=[0, 0, -2, 1]^T$ so that the decoded signal is $\bbc^{*T}_{10,\mathrm{F}} \, [0,0,w,g+2w]^T = g$, as desired. 
Fig.~\ref{fig:example_anc} demonstrates that, through the optimal design of node-variant filters, every node can recover its target signal after three information exchanges with neighbors. Notice that three is also the smallest number of exchanges that can yield perfect recovery since sink nodes 7 and 10 are \emph{three} hops away from their associated source node 3. Moreover, note that the communication protocol induced by the shift operator $\bbS = \bbA$ is extremely simple: each node forwards its current signal to its neighbors and, in turn, receives the sum of their signals.

%%%%%%%%%%% T A B L E %%%%%%%%%
%\begin{table}[t]
%	\centering
%	\small
%	\begin{tabular}{c | c | c c c c}
%		& recover & \multicolumn{4}{c}{observed signal $\bbz^{(t)}$}\\
%		node & signal & $t=0$ & $t=1$ & $t=2$ & $t=3$ \\ \hline
%		1 & $g$	& 0 	& $g$ 	& $g+w$ 	& $5g+2w$\\
%		2 & $w$ 	& 0 	& $g$ 	& $g+w$	& $5g+3w$\\
%		3 & $w$	&$g$	& 0 		& $4g+2w$& $4g+3w$\\
%		4 & $g$	& 0	& $g + w$	& $g$	& $7g+7w$\\
%		5 & $w$	& 0	& $g + w$	& $g+w$	& $5g+8w$\\
%		6 & $g$	&$w$ & 0 		& $2g+4w$& $3g+2w$\\
%		7 & $g$	& 0 	& $w$ 	& 0 		& $2g+5w$\\
%		8 & $w$	& 0 	& $w$ 	& $g+w$	& $4g+7w$\\
%		9 & $w$	& 0	& 0 		& $g+2w$	& $2g+3w$\\
%		10& $g$	& 0 	& 0		& $w$	& $g+2w$\\ \hline
%	\end{tabular}
%	\caption{Signal to recover at each sink node in the graph in Fig.~\ref{fig:example_anc} and observed shifted signals for $t \in \{0, 1, 2, 3\}$.}
%	\label{tab:anc_example}
%\end{table}
%%%%%%%%%%%%%%%%%%%%%%%%%

The following experiments evaluate the ensemble performance by averaging the results of multiple tests. To that end, we consider Erd\H{o}s-R\'{e}nyi graphs with $N=100$,  $p_{edge}=0.1$ and weights drawn from a uniform distribution with support $[0.5, 1.5]$, and $\bbS=\bbA$. We randomly select $S=5$ sources and, to each of these sources, we assign a sink so that $R=5$ and $\bbB_{\ccalS \ccalR}= \bbI$. {The} goal is to evaluate the performance gap when approximating $\bbB_{\ccalS\ccalR}$ vs. $\bbB_{\ccalR}$. Recall that the former assumes that the nodes that are not sources inject a zero input. Alternatively, in the latter each receiver only knows its intended source, so that it must annihilate the signal from all other nodes. Since the size of $\bbB_{\ccalR}$ is $N/S$ times larger than the size of $\bbB_{\ccalS\ccalR}$, the performance of the former is expected to be considerably lower. To corroborate this, Fig.~\ref{fig:sub2_anc} plots the mean error across 1,000 graphs for node-invariant (red) and node-variant (blue) filters when the coefficients are designed to approximate both $\bbB_{\ccalR}$ and $\bbB_{\ccalS\ccalR}$. Denoting by $\bbx$ the $S$-sparse input signal containing the values to be transmitted by the source nodes (drawn from an smGd) and by $\bby = \bbH \bbx$ the filtered signal, the error is defined as $e:=\|\bbE_\ccalR \bby - \bbE_\ccalS \bbx\|_2/\| \bbE_\ccalS \bbx \|_2$, i.e. the normalized difference between the signal injected at the sources and the one recovered at the sinks. Notice that if the sources are known and the relay nodes inject a zero signal, after 7 local exchanges the mean error for node-variant filters is 0. For this same filter degree the error when approximating $\bbB_{\ccalR}$ is 0.93. Eventually, the latter error also vanishes, but filters of degree close to $N\!=\!100$ are needed. By contrast, when node-invariant GFs are used to approximate $\bbB_{\ccalR}$, the error improvement associated with increasing filter degree for the selected interval ($0\leq L-1 \leq 9$) is negligible. Finally, by comparing the plots for node-variant and node-invariant GFs when the coefficients are designed to approximate  $\bbB_{\ccalS\ccalR}$, we observe that node-variant GFs achieve zero error after a few interactions, while node-invariant GFs exhibit a slow reduction of the error with the filter degree.

Correlation among the injected signals at source nodes can be leveraged to reduce the error at the receivers. To illustrate this, for different values of $\rho \in \{0, 0.2, \ldots, 1\}$, we build an $S \times S$ covariance matrix $\bbR_{\bbx, \rho}$ defined as $\bbR_{\bbx, \rho}^{1/2} := \bbI + \rho (\mathbf{1} \mathbf{1}^T - \bbI) + 0.1 \rho \bbZ$ where the elements in the \emph{symmetric} matrix $\bbZ$ are drawn from an smGd. In this way, the correlation between injected signals increases with $\rho$ and, in each realization, is corrupted by additive zero-mean random noise. In Fig.~\ref{fig:sub3_anc} we plot the mean error across 1,000 graphs as a function of the node-variant filter degree parametrized by $\rho$. Notice first that when $\rho=0$, the values injected at different sources are uncorrelated and we recover the blue plot in Fig.~\ref{fig:sub2_anc}. More interestingly, as $\rho$ increases, the achieved error for a given filter degree markedly decreases. 
%For example, after 3 exchanges the mean error for uncorrelated sources is 0.68 whereas for $\rho=1$ this error is reduced to 0.12. 
The reason for this is that correlation allows sink nodes to form accurate estimates of their objective source signals based on \emph{all} injected signals.

%%%%%%%%%%%%%%%   F   I   G   U   R    E   :   R    A    N    D        N   E   T   W   O   R   K   %%%%%%%%%%%%%%%%%%%%%
\begin{figure*}[t]
		\centering	
		\begin{subfigure}{.245\textwidth}
		\centering
        \includegraphics[width=\linewidth, height = 0.685\linewidth]{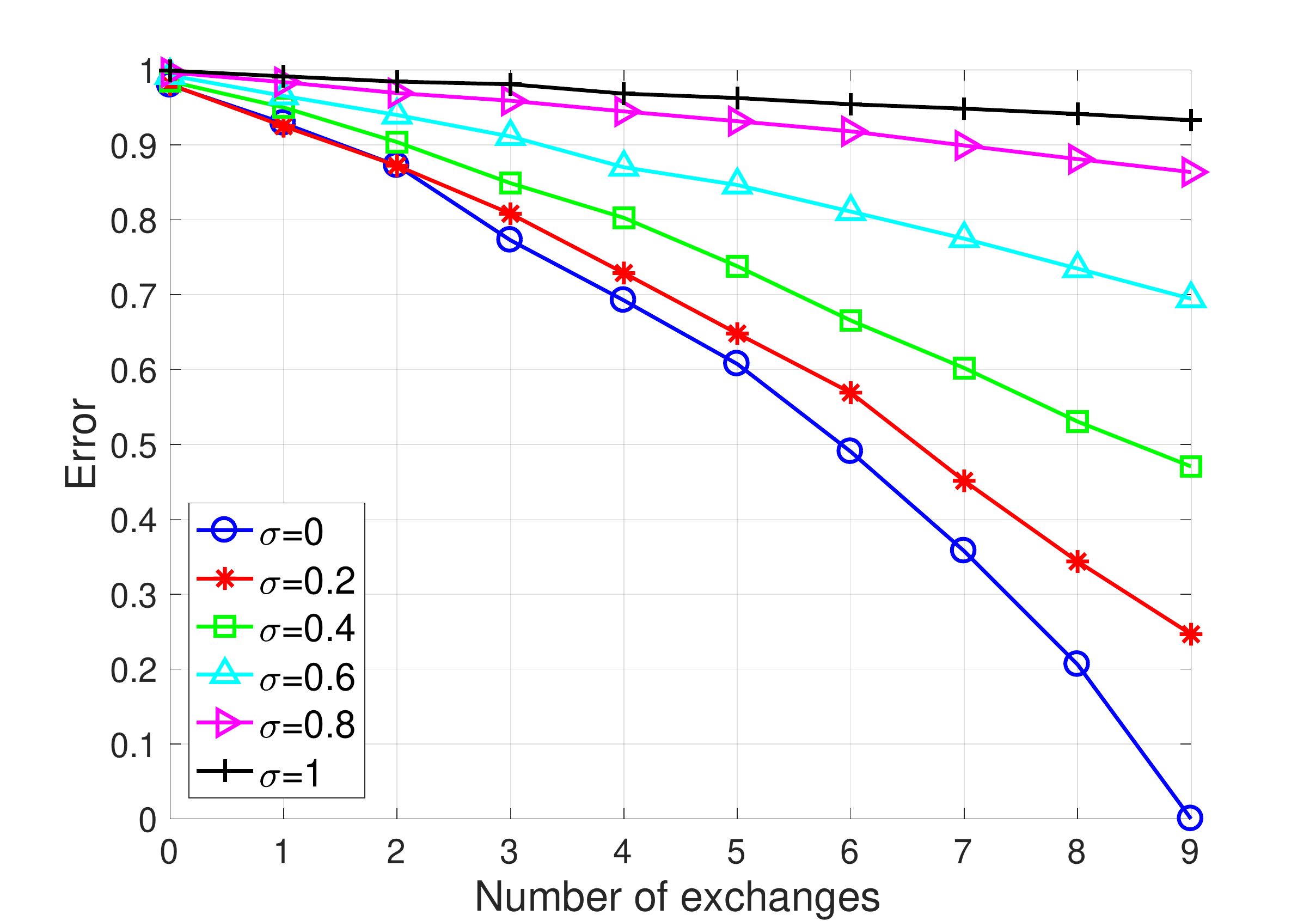}
		\caption{{Node invariant - noisy $\bbV$}}
				\label{fig:extra_sub1}
      \end{subfigure}
      \begin{subfigure}{.245\textwidth}
      	\centering
      	  \includegraphics[width=\linewidth, height = 0.685\linewidth]{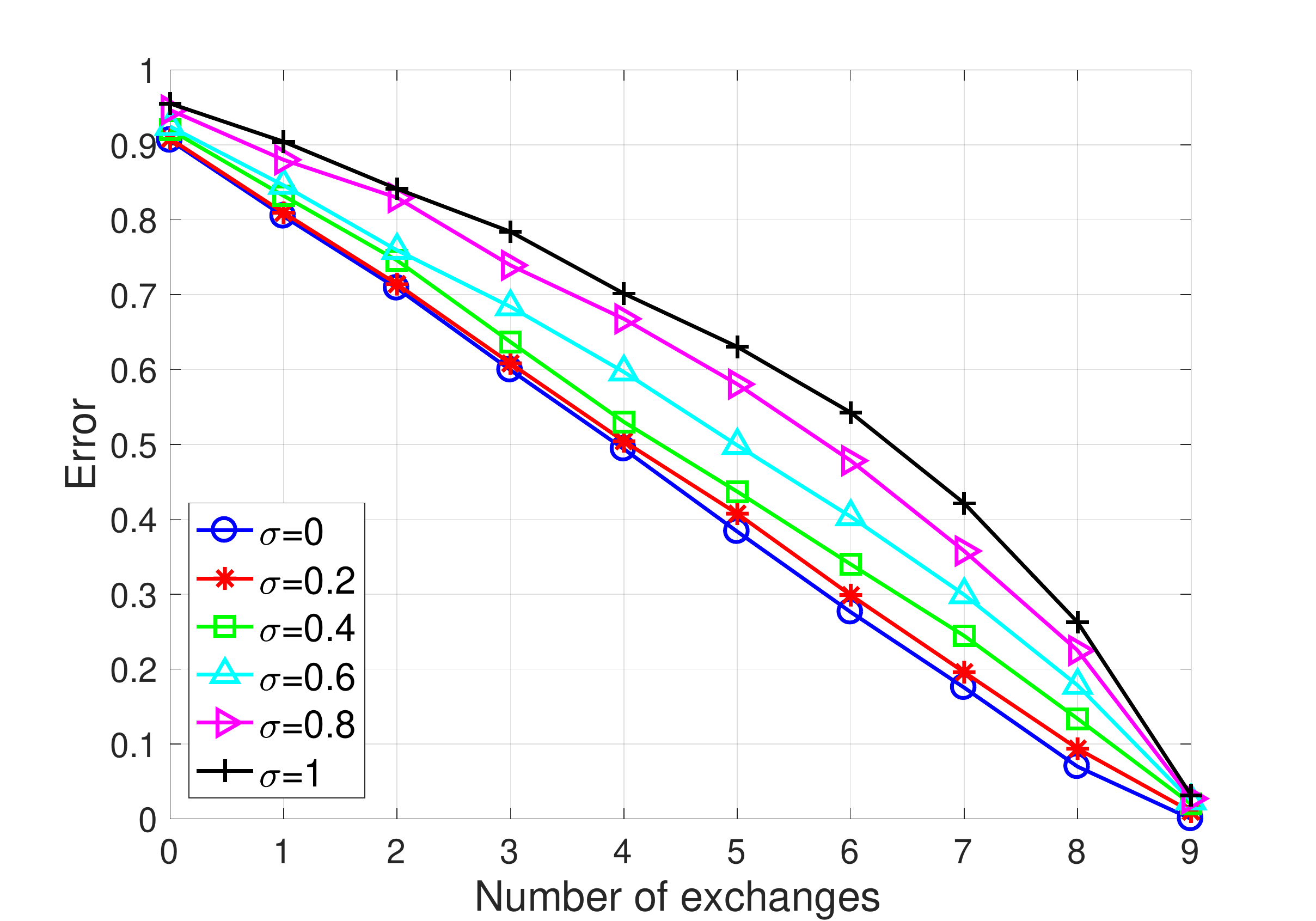}
		\caption{{Node variant - noisy $\bbV$}}
		\label{fig:extra_sub4}
      \end{subfigure}
      \begin{subfigure}{.245\textwidth}
      	\centering
      	  \includegraphics[width=\linewidth, height = 0.685\linewidth]{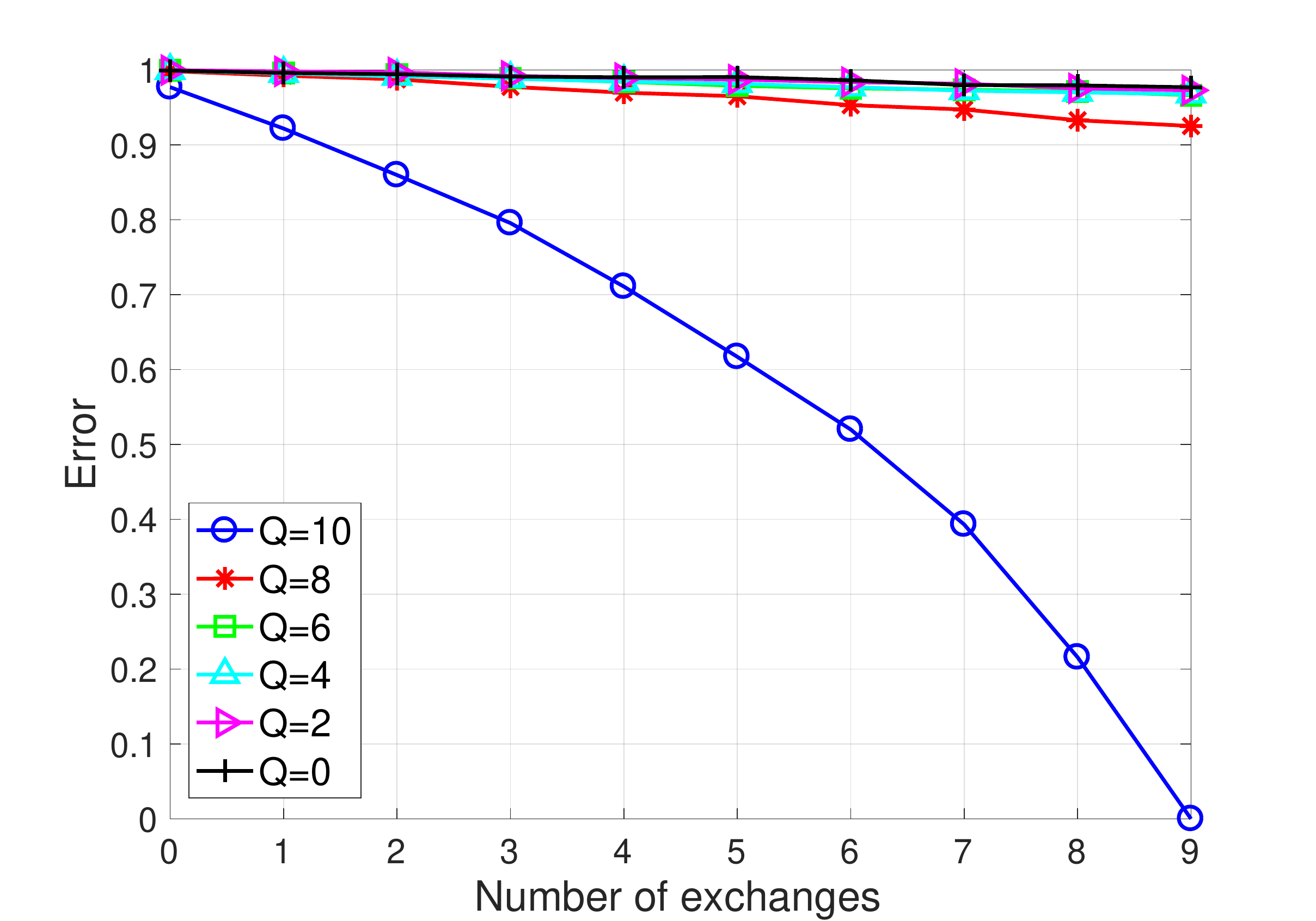}
		\caption{{Node invariant - incomplete $\bbV$}}
				\label{fig:extra_sub5}
      \end{subfigure}
      \begin{subfigure}{.245\textwidth}
      	\centering
      	  \includegraphics[width=\linewidth, height = 0.685\linewidth]{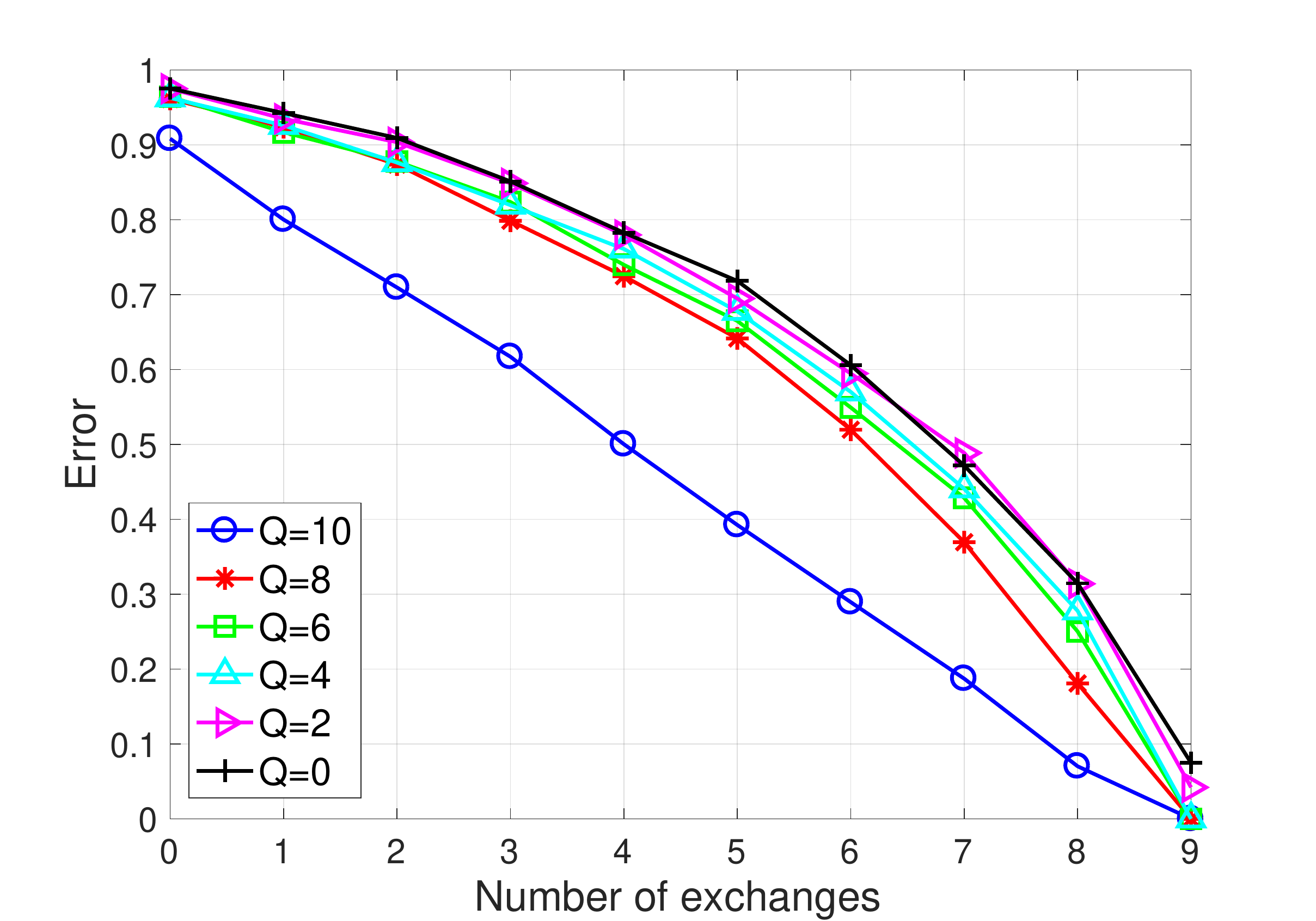}
		\caption{{Node variant - incomplete $\bbV$}}
				\label{fig:extra_sub6}
      \end{subfigure}
    \caption{{Approximation errors as a function of the filter degree based on 1,000 realizations and small-world graphs. 
(a) Spectrum mismatch for a node-invariant GF with $N=10$ and symmetric shift. Curves represent median recovery error parametrized by the eigenbasis perturbation magnitude $\sigma$. 
(b) Same as in (a) but for node-variant GFs.
(c) Same as in (a) but parametrized by $Q$, the number of shared eigenvectors between $\bbS$ and $\bbB$.
(d) Same as in (c) but for node-variant GFs.}}
  \label{fig_extra_simus}
\end{figure*}
%%%%%%%%%%%%%%%%%%%%%%%%%%%%%%%%%%%%%%%%%%%%%%%%%%%%%%%%%%%%%%%%%%%

{
%%%%%%%%%%%%%%%%%%%%%%%%%%%%%%%%%%%%%%%%%%
\subsection{Optimal graph-shift operator design}\label{Ss:optimal_shift_design}
%%%%%%%%%%%%%%%%%%%%%%%%%%%%%%%%%%%%%%%%%%
	
While the previous experiments investigated the design of the filter coefficients for a given shift, the focus of this section is on designing the weights of the shift itself. Specifically, we test the results in Section~\ref{Ss:Designing_the_shift} regarding the implementation of rank-one transformations $\bbB$ using node-invariant GFs.
	
To perform these tests, we generate Erd\H{o}s-R\'enyi graphs $\ccalG$ with $N=10$ nodes and edge probability $p_{\mathrm{edge}}=0.3$, and rank-one transformations $\bbB = \bba \bbb^T$ where $\bba$ and $\bbb$ are drawn from an smGd. For each graph we consider three different shift operators: i) the \emph{original shift} $\bbS_1$, where we assign random weights to every edge present in the graph and its diagonal; ii) the \emph{re-weighted tree} $\bbS_2$, where we randomly select a spanning tree of $\ccalG$ and then set the weights as dictated by \eqref{E:rankoneshif_def}; and iii) the \emph{re-weighted shift} $\bbS_3$ where we use \eqref{E:rankoneshif_def} to select weights for \emph{all} the edges in $\ccalG$, as opposed to just those forming a spanning tree. 
We test the performance of approximating $\bbB$ as node-invariant filters $\bbH(\bbc, \bbS_m)$ for $m \in \{1, 2, 3\}$ and for varying filter degrees. Specifically, for each filter degree or number of exchanges $K \in \{0, 1, \ldots, 9\}$ we obtain the optimal GF coefficients $\bbc_m^K$ for shift $\bbS_m$ by solving \eqref{E:optimal_filter_coefficients_case_2}, where $\bbR_{\bbx}=\bbI$. We then measure the error as $\| \bbB - \bbH(\bbc_m^K, \bbS_m) \|_{\mathrm{F}}/\| \bbB \|_{\mathrm{F}}$ and plot its average across $1,000$ realizations in Fig.~\ref{fig_optimal_shift_design}.
	
GFs based on the original shift $\bbS_1$ approximate poorly the rank-one transformations, achieving a large error of around $0.95$ even for filters of degree $9$. This is not surprising since, in general, $\bbS_1$ and $\bbB$ will not be simultaneously diagonalizable (cf. Proposition~\ref{P:Perfect_Impl_NodeInvarFilter}) and node-invariant filters are very sensitive to spectral misalignments (cf. Section~\ref{Ss:NumExper_SpectralRob}). The tree-based shift $\bbS_2$ can perfectly implement $\bbB$ for a large enough degree, confirming the result shown in Proposition~\ref{P:shift_rankone}. Equally interesting, the full-graph shift $\bbS_3$ also attains zero error. Recall that, for simplicity, the constructive proof of Proposition~\ref{P:shift_rankone} that guarantees zero-error was based on spanning trees. In line with the discussion following Proposition~\ref{P:shift_rankone}, the simulations support the presumption that the construction is also valid for more general graphs. Note finally that, when comparing the performance of $\bbS_2$ and $\bbS_3$, the latter tends to outperform the tree-based design. This is somehow expected since the larger number of links in $\bbS_3$ facilitates the quick propagation of information across the graph. Indeed, as $K$ increases and approaches the diameter of the tree both designs perform similarly.
}

\vspace{.3cm}

%%%%%%%%%%%%%%%%%%%%%%%%%%%%%%%%%%%%%%%%%%
\subsection{Spectral robustness of GF implementations}\label{Ss:NumExper_SpectralRob}
%%%%%%%%%%%%%%%%%%%%%%%%%%%%%%%%%%%%%%%%%%

A limitation of \emph{node-invariant} GFs is that perfect implementation of a transformation $\bbB$ requires $\bbS$ and $\bbB$ to share the whole set of eigenvectors. In this section, we run numerical simulations to assess: i) how the implementation degrades as the eigenvectors of $\bbS$ and $\bbB$ are further apart; and ii) whether the implementation using \emph{node-variant} filters is robust to this dissimilarity.  For each experiment, we generate 1,000 unweighted and symmetric {small-world graphs with $N=10$, where each node is originally connected in average to four neighbors, and rewiring probability of $0.2$.} We set $\bbS=\bbA$, and define a signal $\bbx$ drawn from an smGd on each graph. For a given degree, the filter coefficients are obtained using the optimal designs in \eqref{E:optimal_filter_coefficients_case_2} and \eqref{E:optimal_filter_coefficients_case_2_var}.

Two procedures are considered to construct linear transformations $\bbB$ whose eigenvectors have varying degrees of similarity with those of $\bbS = \bbV \bbLambda \bbV^{H}$. In the first one, we generate perturbed eigenbases from $\bbV$ by adding elementwise zero-mean Gaussian noise of varying power. Specifically, we build $\bbB = \bbV_\bbB^\sigma \diag(\bbbeta) (\bbV_\bbB^\sigma)^{-1}$ with the eigenvalues $\bbbeta$ drawn from an smGd and the eigenvectors $\bbV_\bbB^\sigma$ obtained by normalizing the columns of $\bbV + \sigma \bbZ \circ \bbV$, where the elements in the $N \times N$ matrix $\bbZ$ also follow an smGd and $\circ$ denotes the Hadamard matrix product. 
%In this way, the larger the scalar $\sigma$, the larger the perturbation applied on $\bbV$ to obtain $\bbV_\bbB^\sigma$. 
Defining the error as $e:=\|\bbH\bbx - \bbB \bbx\|_2 / \| \bbB \bbx\|_2$, in Fig.~\ref{fig:extra_sub1} we present the median error achieved by node-invariant filters when approximating $\bbB$. Each curve corresponds to a different value of $\sigma$. For the case where $\sigma = 0$, $\bbS$ and $\bbB$ are simultaneously diagonalizable and, hence, perfect implementation is achievable after $N-1$ exchanges among neighbors. As expected the error increases with $\sigma$, e.g., when $\sigma=0.2$ the {best achievable error is around 0.25 for filters of degree 9}. Part of this insurmountable error comes from the fact that $\bbB$ is, in general, asymmetric while node-invariant filters $\bbH$ inherit the symmetry from $\bbS$. By contrast, node-variant filters can successfully implement $\bbB$ for all values of $\sigma$; see Fig.~\ref{fig:extra_sub4}. Even though there is a slight dependence with $\sigma$, all curves demonstrate that the error decreases with the number of exchanges until almost perfect implementation is achieved for $K\!=\!N\!-\!1$. The second procedure keeps only a subset of the eigenvectors of the shift. More specifically, we build linear operators of the form $\bbB = \bbV_\bbB^Q \diag(\bbbeta) (\bbV_\bbB^Q)^{-1}$ where $Q$ of the columns of matrix $\bbV_\bbB^Q$ are chosen randomly without replacement from $\bbV$. The remaining $N-Q$ columns correspond to random linearly independent, non-orthogonal, vectors. In this way, the resulting $\bbV_\bbB^Q$ is not unitary and $\bbB$ is non-symmetric (symmetry will require setting the $N-Q$ vectors as an orthonormal basis of the orthogonal complement of the space spanned by the $Q$ columns selected from $\bbV$). In Fig.~\ref{fig:extra_sub5} we present the median error attained by node-invariant filters parametrized by $Q$. As expected, when $Q=10$, perfect implementation is achieved since $\bbB$ and $\bbS$ are simultaneously diagonalizable. As $Q$ decreases, a noticeable detrimental effect is observed. The reason being that node-invariant GFs with symmetric shifts can give rise only to symmetric transformations, and $\bbB$ here is non-symmetric. Indeed, if the remaining $N-Q$ columns of $\bbV_\bbB^Q$ are chosen such that $\bbV_\bbB^Q$ is unitary (and, hence, $\bbB$ is symmetric), the approximation error for intermediate values of $K$ is considerably smaller (not shown here).
%In the limit, when $Q=0$, the transformation $\bbB$ to implement is an arbitrary \emph{symmetric} operator and the performance of node-invariant filters is unsatisfactory with a median error of 0.88. 
By contrast, node-variant filters show imperviousness to the degree of overlap $Q$ between the eigenbases of $\bbS$ and $\bbB$; see Fig.~\ref{fig:extra_sub6}. 
%Moreover, the strict monotonicity with respect to $\sigma$ regardless of the value of $L-1$ observed in Fig.~\ref{fig:extra_sub4} is not present here. In fact, when $L-1$ is small, slightly smaller errors are attained for low values of $Q$, while the opposite is true when $L-1$ is large. 

%Finally, we consider the scenario where, in forming $\bbV_\bbB^Q$, we keep $Q$ eigenvectors from $\bbV$ as in the previous case but the remaining $N-Q$ eigenvectors are random but not necessarily orthogonal. Consequently, the obtained transformation $\bbB$ is, in general, asymmetric. This drastically affects the implementation performance using GFs, especially for the node-invariant case where the filter obtained from a symmetric $\bbS$ must be symmetric. In Fig.~\ref{fig:extra_sub3}, we observe that just by reducing $Q=10$ (total overlap) to $Q=8$, the minimum error ($K=9$) grows from 0 to 0.9, making node-invariant filters unsuitable for this class of implementations. Asymmetry also affects the performance of node-variant filters (see Fig.~\ref{fig:extra_sub6}), especially for intermediate degrees.

In a nutshell, the performance of \textit{node-invariant} GFs depends noticeably on the similarity between the eigenvectors of $\bbS$ and $\bbB$. The approximation error grows gradually with the distance between eigenbases, and larger errors are observed when approximating asymmetric operators with symmetric shifts.  {This behavior was also observed in \cite{ssamar_distfilters_allerton15} for Erd\H{o}s-R\'{e}nyi graphs.} By contrast, \textit{node-variant} GFs are robust to the spectral differences between $\bbS$ and $\bbB$. A theoretical characterization of the described robustness as well as an analysis of its dependence on topological features of the underlying graph are interesting research directions and are left as future work.

%%%%%%%%%%%%%%%%%%%%%%%%%%%%%%%%%%%%%%%%%%
\section{Conclusions}\label{S:Concl}
%%%%%%%%%%%%%%%%%%%%%%%%%%%%%%%%%%%%%%%%%%

The optimal design of GFs for (potentially distributed) implementation of generic linear operators was investigated. Our first contribution was the introduction of two types of node-variant GFs, which generalize the existing definition of (node-invariant) GFs by allowing each node to use a different set of filter coefficients. Then, for both node-invariant and node-variant filters, we stated conditions for perfect implementation of linear operators. When these conditions were not met, we provided optimal filter-coefficient designs for the minimization of different error metrics. We also addressed the optimal design of the shift operator and provided a systematic way to define a shift that allows the implementation of any rank-one operator as a GF. The practical relevance of our approach for {\textit{distributed}} setups was emphasized by particularizing our results to the design of GFs able to implement finite-time consensus and analog network coding, two problems commonly studied in network settings. 
%Finally, we relied on numerical results to compare the performance of node-invariant and node-variant approximations and related it to spectral features of the linear operators to implement.

%%%%%%%%%%%%%%%%%%%%%%%%%%%%
\bibliographystyle{IEEEtran}
%% argument is your BibTeX string definitions and bibliography database(s)
%
\bibliography{citations}

\begin{thebibliography}{10}
\providecommand{\url}[1]{#1}
\csname url@samestyle\endcsname
\providecommand{\newblock}{\relax}
\providecommand{\bibinfo}[2]{#2}
\providecommand{\BIBentrySTDinterwordspacing}{\spaceskip=0pt\relax}
\providecommand{\BIBentryALTinterwordstretchfactor}{4}
\providecommand{\BIBentryALTinterwordspacing}{\spaceskip=\fontdimen2\font plus
\BIBentryALTinterwordstretchfactor\fontdimen3\font minus
  \fontdimen4\font\relax}
\providecommand{\BIBforeignlanguage}[2]{{%
\expandafter\ifx\csname l@#1\endcsname\relax
\typeout{** WARNING: IEEEtran.bst: No hyphenation pattern has been}%
\typeout{** loaded for the language `#1'. Using the pattern for}%
\typeout{** the default language instead.}%
\else
\language=\csname l@#1\endcsname
\fi
#2}}
\providecommand{\BIBdecl}{\relax}
\BIBdecl

\bibitem{ssamar_distfilters_allerton15}
S.~Segarra, A.~G. Marques, and A.~Ribeiro, ``Distributed implementation of
  linear network operators using graph filters,'' in \emph{53rd Allerton Conf.
  on Commun. Control and Computing}, Sept. 2015, pp. 1406--1413.

\bibitem{ssamar_distfilters_icassp16}
------, ``Linear network operators using node-variant graph filters,'' in
  \emph{IEEE Intl. Conf. Acoust., Speech and Signal Process. (ICASSP)}, Mar.
  2016, pp. 4850--4854.

\bibitem{EmergingFieldGSP}
D.~Shuman, S.~Narang, P.~Frossard, A.~Ortega, and P.~Vandergheynst, ``The
  emerging field of signal processing on graphs: Extending high-dimensional
  data analysis to networks and other irregular domains,'' \emph{IEEE Signal
  Process. Mag.}, vol.~30, no.~3, pp. 83--98, May 2013.

\bibitem{SandryMouraSPG_TSP13}
A.~Sandryhaila and J.~Moura, ``Discrete signal processing on graphs,''
  \emph{IEEE Trans. Signal Process.}, vol.~61, no.~7, pp. 1644--1656, Apr.
  2013.

\bibitem{SamplingKovacevic_without_Moura_15}
S.~Chen, R.~Varma, A.~Sandryhaila, and J.~Kova{\v{c}}evi{\'c}, ``Discrete
  signal processing on graphs: Sampling theory,'' \emph{IEEE Trans. Signal
  Process.}, vol.~63, no.~24, pp. 6510 -- 6523, Dec. 2015.

\bibitem{marques2016stationaryTSP16}
A.~G. Marques, S.~Segarra, G.~Leus, and A.~Ribeiro, ``Stationary graph
  processes and spectral estimation,'' \emph{arXiv preprint arXiv:1603.04667},
  2016.

\bibitem{isufi2016separable}
E.~Isufi, A.~Loukas, A.~Simonetto, and G.~Leus, ``Separable autoregressive
  moving average graph-temporal filters,'' in \emph{European Signal Process.
  Conf. (EUSIPCO)}, Budapest, Hungary, Aug. 29 - Sept. 2 2016.

\bibitem{isufi2017autoregressive}
------, ``Autoregressive moving average graph filtering,'' \emph{IEEE
  Transactions on Signal Processing}, vol.~65, no.~2, pp. 274--288, 2017.

\bibitem{SandryMouraSPG_TSP14Freq}
A.~Sandryhaila and J.~Moura, ``Discrete signal processing on graphs: Frequency
  analysis,'' \emph{IEEE Trans. Signal Process.}, vol.~62, no.~12, pp.
  3042--3054, June 2014.

\bibitem{segarra2016topologyid}
S.~Segarra, A.~G. Marques, G.~Mateos, and A.~Ribeiro, ``Network topology
  inference from spectral templates,'' \emph{arXiv preprint
  arXiv:1608.03008v1}, 2016.

\bibitem{bertsekas1989parallelBook}
D.~P. Bertsekas and J.~N. Tsitsiklis, \emph{Parallel and Distributed
  Computation: Numerical Methods}.\hskip 1em plus 0.5em minus 0.4em\relax
  Prentice-Hall, Inc., 1989.

\bibitem{XiaoBoyd2004}
L.~Xiao and S.~Boyd, ``Fast linear iterations for distributed averaging,''
  \emph{Systems \& Control Lett.}, vol.~53, no.~1, pp. 65 -- 78, 2004.

\bibitem{kar2013distributed}
S.~Kar, J.~Moura, and H.~Poor, ``Distributed linear parameter estimation:
  asymptotically efficient adaptive strategies,'' \emph{{SIAM} Journal on
  Control and Optimization}, vol.~51, no.~3, pp. 2200--2229, May 2013.

\bibitem{schizas2008consensus}
I.~D. Schizas, A.~Ribeiro, and G.~B. Giannakis, ``Consensus in ad-hoc {WSN}s
  with noisy links - {P}art {I}: Distributed estimation of deterministic
  signals,'' \emph{IEEE Trans. Signal Process.}, vol.~56, no.~1, pp. 350--364,
  Jan. 2008.

\bibitem{AliConsensus08}
A.~Tahbaz-Salehi and A.~Jadbabaie, ``A necessary and sufficient condition for
  consensus over random networks,'' \emph{IEEE Trans. Auto. Control}, vol.~53,
  no.~3, pp. 791--795, Apr. 2008.

\bibitem{mateos2010dlasso}
G.~Mateos, J.-A. Bazerque, and G.~B. Giannakis, ``Distributed sparse linear
  regression,'' \emph{IEEE Trans. Signal Process.}, vol.~58, no.~10, pp.
  5262--5276, Oct. 2010.

\bibitem{IntroSPMag_2013_Nets}
A.~Sayed, S.~Barbarossa, S.~Theodoridis, and I.~Yamada, ``Adaptation and
  learning over complex networks [from the guest editors],'' \emph{IEEE Signal
  Process. Mag.}, vol.~30, no.~3, pp. 14--15, May 2013.

\bibitem{BarbarossaIcassp09}
S.~Barbarossa, G.~Scutari, and T.~Battisti, ``Distributed signal subspace
  projection algorithms with maximum convergence rate for sensor networks with
  topological constraints,'' in \emph{IEEE Intl. Conf. Acoust., Speech and
  Signal Process. (ICASSP)}, Apr. 2009, pp. 2893--2896.

\bibitem{ShumFrossard_ChebyshevDist2011}
D.~I. Shuman, P.~Vandergheynst, and P.~Frossard, ``Distributed signal
  processing via {Chebyshev} polynomial approximation,'' \emph{arXiv preprint
  arXiv:1111.5239}, 2011.

\bibitem{wang2015local}
X.~Wang, P.~Liu, and Y.~Gu, ``Local-set-based graph signal reconstruction,''
  \emph{IEEE Trans. Signal Process.}, vol.~63, no.~9, pp. 2432--2444, Sept.
  2015.

\bibitem{Kovac_Dist_InpaintingICASPP15}
S.~Chen, A.~Sandryhaila, and J.~Kovacevic, ``Distributed algorithm for graph
  signal inpainting,'' in \emph{IEEE Intl. Conf. Acoust., Speech and Signal
  Process. (ICASSP)}, Apr. 2015, pp. 3731 -- 3735.

\bibitem{sandryhaila2014finite}
A.~Sandryhaila, S.~Kar, and J.~Moura, ``Finite-time distributed consensus
  through graph filters,'' in \emph{IEEE Intl. Conf. Acoust., Speech and Signal
  Process. (ICASSP)}, May 2014, pp. 1080--1084.

\bibitem{SuccNullingEigenv}
S.~Safavi and U.~Khan, ``Revisiting finite-time distributed algorithms via
  successive nulling of eigenvalues,'' \emph{IEEE Signal Process. Lett.},
  vol.~22, no.~1, pp. 54--57, Jan. 2015.

\bibitem{EUSIPCO_our_interp_2015}
S.~Segarra, A.~G. Marques, G.~Leus, and A.~Ribeiro, ``Interpolation of graph
  signals using shift-invariant graph filters,'' in \emph{European Signal
  Process. Conf. (EUSIPCO)}, Aug. 2015, pp. 210--214.

\bibitem{segarra2015reconstruction}
------, ``Reconstruction of graph signals through percolation from seeding
  nodes,'' \emph{IEEE Trans. Signal Process.}, vol.~64, no.~16, pp. 4363 --
  4378, Aug. 2016.

\bibitem{finiteconsensusKibangou11}
A.~Y. Kibangou, ``Finite-time average consensus based protocol for distributed
  estimation over {AWGN} channels,'' in \emph{IEEE Conf. Decis. Contr. (CDC)},
  Dec. 2011, pp. 5595--5600.

\bibitem{Kattietal07ANC}
S.~Katti, S.~Gollakota, and D.~Katabi, ``Embracing wireless interference:
  Analog network coding,'' \emph{SIGCOMM Comput. Commun. Rev.}, vol.~37, no.~4,
  pp. 397--408, Oct 2007.

\bibitem{Zhangetal06ANC}
S.~Zhang, S.~C. Liew, and P.~P. Lam, ``Physical-layer network coding,'' in
  \emph{Intl. Conf. Mobile Comp. and Netw. (MobiCom)}, 2006, pp. 358--365.

\bibitem{godsil2001algebraic}
C.~Godsil and G.~Royle, \emph{Algebraic Graph Theory}.\hskip 1em plus 0.5em
  minus 0.4em\relax Springer-Verlag, Graduate Texts in Mathematics, 2001, vol.
  207.

\bibitem{pukelsheim1993optimal}
F.~Pukelsheim, \emph{Optimal Design of Experiments}.\hskip 1em plus 0.5em minus
  0.4em\relax SIAM, 1993, vol.~50.

\bibitem{Golub_MatComp_Book}
G.~Golub and C.~Van~Loan, \emph{Matrix Computations (3rd Ed.)}.\hskip 1em plus
  0.5em minus 0.4em\relax Johns Hopkins Univ. Press, 1996.

\bibitem{overton1988minimizing}
M.~L. Overton, ``On minimizing the maximum eigenvalue of a symmetric matrix,''
  \emph{SIAM Journal on Matrix Analysis and Applications}, vol.~9, no.~2, pp.
  256--268, 1988.

\bibitem{boyd1993method}
S.~Boyd and L.~El~Ghaoui, ``Method of centers for minimizing generalized
  eigenvalues,'' \emph{Linear Algebra and its Applications}, vol. 188, pp.
  63--111, 1993.

\bibitem{ho2008network}
T.~Ho and D.~Lun, \emph{Network Coding: An Introduction}.\hskip 1em plus 0.5em
  minus 0.4em\relax Cambridge University Press, 2008.

\bibitem{Ahlswede00Network}
R.~Ahlswede, N.~Cai, S.-Y. Li, and R.~Yeung, ``Network information flow,''
  \emph{IEEE Trans. Inf. Theory}, vol.~46, no.~4, pp. 1204--1216, Jul 2000.

\bibitem{LinNetCoding2003}
S.-Y. Li, R.~Yeung, and N.~Cai, ``Linear network coding,'' \emph{IEEE Trans.
  Inf. Theory}, vol.~49, no.~2, pp. 371--381, Feb. 2003.

\bibitem{Nakano_etal12}
T.~Nakano, M.~Moore, F.~Wei, A.~Vasilakos, and J.~Shuai, ``Molecular
  communication and networking: Opportunities and challenges,'' \emph{IEEE
  Trans. Nanobiosci.}, vol.~11, no.~2, pp. 135--148, June 2012.

\bibitem{Kuran201086}
M.~S. žKuran, H.~B. Yilmaz, T.~Tugcu, and B.~O–zerman, ``Energy model for
  communication via diffusion in nanonetworks,'' \emph{Nano Commun. Networks},
  vol.~1, no.~2, pp. 86 -- 95, 2010.

\bibitem{kolaczyk2009book}
E.~D. Kolaczyk, \emph{Statistical Analysis of Network Data: Methods and
  Models}.\hskip 1em plus 0.5em minus 0.4em\relax Springer, 2009.

\end{thebibliography}

\end{document}